\documentclass{ws-mplb}
\renewcommand{\theequation}{\arabic{section}.\arabic{equation}} 

\newcommand{\be}{\begin{eqnarray}}
\newcommand{\ee}{\end{eqnarray}}
\newcommand{\nn}{\nonumber\\}
\newcommand{\la}{\langle}
\newcommand{\ra}{\rangle}
\newcommand{\da}{\downarrow}
\newcommand{\ua}{\uparrow}

\begin{document}

\markboth{I.Ichinose and T.Matsui}{Lattice gauge theory for  
condensed matter physics}

%
\catchline{}{}{}{}{}
%

\title{LATTICE GAUGE THEORY FOR CONDENSED MATTER PHYSICS:
FERROMAGNETIC SUPERCONDUCTIVITY AS ITS EXAMPLE
}

\author{\footnotesize IKUO ICHINOSE}

\address{Department of Applied Physics, 
Nagoya Institute of Technology, \\
Nagoya, 466-8555, Japan
}

\author{TETSUO MATSUI}

\address{Department of Physics, Kinki University,\\
Higashi-Osaka, 577-8502, Japan}

\maketitle

\begin{history}
\received{(Day Month Year)}
\revised{(Day Month Year)}
\end{history}

\begin{abstract}
Recent theoretical studies of various strongly-correlated systems 
in condensed matter physics
reveal that the lattice gauge theory(LGT) developed in high-energy physics
is quite a useful tool to understand physics of these systems. 
Knowledges of LGT are to become a necessary item even for condensed matter
physicists. 
In the first part of this paper, we present a  concise review of 
LGT for the reader who wants to understand its basics for the first time.
For illustration, we choose the abelian Higgs model, 
a typical and quite useful LGT, which is the lattice verison of 
the Ginzburg-Landau model interacting with a U(1) gauge field (vector potential). 
In the second part, we present an account of the recent 
progress in the study of  ferromagnetic superconductivity (SC)
as an example of application of LGT to topics in condensed matter physics, .
As the ferromagnetism (FM) and SC are competing orders with each other,
large fluctuations are expected to take place and therefore nonperturbative
methods are required for theoretical investigation.
After we introduce a LGT describing the FMSC, we study
its phase diagram and topological excitations (vortices of Cooper pairs) 
by Monte-Carlo simulations. 

\end{abstract}


\section{Introduction}
\setcounter{equation}{0}

A strongly-correlated system is a set of either fermions or bosons
which interact strongly each other. Such  a system  may 
exhibit collective behaviors that cannot be expected from 
the behaviors of each single particle, and 
has been the subject of great interest in various fields of 
condensed matter physics.
For example, the $t-J$ model, which is a typical 
model of electrons interacting via strong Coulomb repulsion,
is taken as a standard model of high-temperature 
superconductors\cite{lee}.
Another example is the Bose Hubbard model\cite{bh}
and the related bosonic $t-J$ model\cite{btj}, which are to be used
to study cold bosonic atoms put on an optical lattice\cite{cba} 
and tuned to have large interaction parameters.
In the last two decades, theoretical approach using gauge theory
 and the associated concepts
has proved itself a convincing way to understand essentials of their physics
\cite{lgt1,lgt2}.   
 
The gauge theory put on a spatial lattice (and even on the imaginary 
discrete time in the path-integral formalism) is known as 
lattice gauge theory (LGT), which is introduced by Wilson
in 1974 as a model of quark confinement\cite{wilson}. Since then,
it has contributed to increase our knowledges 
not only of quantum chromodynamics (QCD) itself, but also of 
nonperturbative aspects of general quantum field theory\cite{lgtrev}.
The reason to consider a lattice instead of the usual continuum space is to
reduce the degrees of freedom from uncountable infinity
to countable infinity, which allows us a concrete definition
of the field theory  in the nonperturbative region and
also ways to examine it such as the strong-coupling (high-temperature) 
expansion.
Concerning to the global phase structure, LGT shows, in addition to
the ordinary Coulomb phase and Higgs phase 
of gauge dynamics (they are sometimes called deconfinement phases),
the existence of so called confinement phase. This phase was crucial
to explain confinement of quarks\cite{wilson}, one of the longstanding problem
 in high-energy physics, because, in the confinement phase,
 the potential energy $V(r)$ between a pair of static quark and anti-quark 
 separated by the distance $r$ increases linearly in $r$
 and costs infinite energy to separate them ($r\to \infty$). 
These three phases are classified by the magnitude of fluctuations 
of gauge field (See Table I below).
The gauge field put on the lattice can be treated as periodic variables
(one calls this case a compact gauge field; see Sect.2.1), and
the gauge-field configurations reflecting this periodicity may 
involve large spatial variations beyond the conventional perturbative ones. 
Monopoles are a typical and important
example of such configurations (monopoles in the continuum
gauge theory cost infinite energy and so not possible).
In the confinement phase, such monopoles condense\cite{thooft}, 
and the resulting strong fluctuations of gauge field are argued to
generate squeezed one-dimensional line of electric flux (variables
conjugate the the gauge field itself). 
Then, in the confinement phase, such an electric flux  gives rise to
the linearly-rising confining potential. 
This phenomenon of formation of electric fluxes is
sometimes called the ``dual" Meissner effect in contrast  
to the Meissner effect in ordinary superconductors where
magnetic field is squeezed into magnetic fluxes due to Cooper-pair
condensation.
One way to recognize the 
relevance of monopoles among other possible (topological) excitations 
is to perform the duality transformation\cite{savit},
which rewrites the system as an ensemble of monopoles or monopole loops.
\vspace{-0.3cm}
\hspace{0.5cm}
\begin{center}
\begin{tabular}{|c|c|c|}
\hline
hase & fluctuation of gauge field &
potential energy $V(r)$  
\\  
\hline
 Higgs & small & $\propto \displaystyle{\frac{\exp(-mr)}{r}}$ \\ \hline
 Coulomb & medium & $\propto \displaystyle{\frac{1}{r}}$ \\ \hline
 Confinement & large & $\propto r$ \\ 
\hline
\end{tabular}  \\
{\small Table 1. Three phases of gauge dynamics: $V(r)$ is the potential energy stored 
between a pair of point charges with the opposite signs and separated by 
the distance $r$.} \\
\end{center}

Also, in some restricted fields of condensed matter physics, 
LGT has been applied successfully.
One example is the $t-J$ model of high-T$_c$ superconductivity (SC)\cite{lgt1}.
Here the strong correlations are  implemented by excluding the double occupancy 
state of electrons, the state of two electrons with the opposite
spin directions occupying the same site, from the physical states.
This may be achieved by viewing an electron as a composite of 
fictitious particles, a so-called holon carrying the charge degree of freedom 
of  the electron and a spinon carrying the spin degree of 
freedom\cite{slave}.
By the method of auxiliary field in path-integral, one may introduce 
U(1) gauge field which binds holon and spinon.
Then the possible phases may be classified according to the dynamics
of this U(1) gauge field. In the confinement phase, the relevant quasiparticles are
the original electrons in which holons and spinons are confined.
This corresponds to the state at the overdoped high temperature region.
In the deconfinement Coulomb phase, quasiparticles are holons and spinons in which
charge and spin degrees of freedom behave independently. This phenomenon
is called the charge-spin separation\cite{css} 
and are to describe the anomalous behavior
of the normal state such as the liner-T resistivity. 
Further correspondence between the gauge dynamics and the observed
phases are also satisfactory as discussed in  Refs.\cite{lee,lgt1,css}.
In particular, possibility of experimental observation of spinons is 
discussed in Ref.\cite{lee2}.  

Another application of LGT to relate
the dynamics of auxiliary gauge field and the observed phases is the
(fractional) Hall effect\cite{jain}. Among a couple of parallel
formulations\cite{fqhe}, one may view an electron  
as a composite of a so called bosonic fluxon carrying fictitious 
(Chern-Simons) gauge flux 
and a so called fermionic chargon carrying the electric charge\cite{lgt2}. 
In the deconfinement phase, 
these fluxons and chargons are separated, i.e., particle-flux separation
takes place. If fluxons Bose condense at sufficiently low temperatures, 
the resulting system is  fermionic
chargons in a reduced uniform magnetic field. 
These chargons are nothing but the Jain's composite fermions, and 
the integer Hall effect of
them explain the fractional Hall effect of the original electron system. 

In this paper, we shall see 
yet another example of LGT
approach to condensed matter physics.
Namely, we consider a lattice model of ferromagnetic SC, which exhibits 
 the ferromagnetism (FM) and/or SC, two typical and important
phenomena in condensed matter physics.
We note here that the introduction of a spatial lattice 
in the original LGT in high-energy theory is traced back, as mentioned,  
to the necessity to define a nonperturbative field theory. The lattice 
spacing there should be taken as a running cutoff(scale) parameter in the 
renormalization-group theory\cite{wilsonkogut} and the continuum limit
should be taken carefully.
However, in condensed matter physics, introduction
of a lattice has often a practical support independent of the above
meaning, i.e., the system itself has a lattice structure.
One may naturally regard the lattice of the model as the real lattice
of material in question. In this paper, we shall not touch the continuum limit
in the sense of renormalization group.   

The layout of the paper is as follows:
In Sect. 2, we review LGT.
The reader who is not familiar with this 
subject may acquire necessary knowledge to understand 
the successive sections.
In Sect. 3, we explain the lattice model of 
ferromagnetic superconductivity (FMSC).
It is based on the known Ginzburg-Landau theory defined in the continuum space.
In Sect. 4, we present numerical results of Monte Carlo simulations for  
the lattice model in Sect. 3.
In Sect. 5, we give conclusion.

\section{Introduction to lattice gauge thoery}
\setcounter{equation}{0}

Formulating field-theoretical models by using space(-time) lattice
is quite useful for studying their properties nonperturbatively.
In particular, because of the lattice formulation itself,
the high-temperature expansion (strong-coupling expansion) can
be performed analytically even for the case in which fluctuations of the fields are
very strong. 
Furthermore, numerical studies like the Monte-Carlo simulation
can be applicable for wide range of the lattice field theories.
In this section,  we review LGT consisting of a U(1) gauge field 
and a charged bosonic matter field.


\subsection{Lattice gauge theory in Hamiltonian formalism}

In this subsection, we explain the Hamiltonian formalism of LGT\cite{ks}.
The reasons are two fold: (i) it gives rise to an intuitive
picture of the relevant states in terms of electric field and magnetic field,
and (ii) it plays an important role in 
recent development in quantum simulations
of the dynamical gauge field by using ultra-cold atomic systems
put on an optical lattice.

\begin{figure}[t]
  \begin{center}
      \includegraphics[width=8cm]{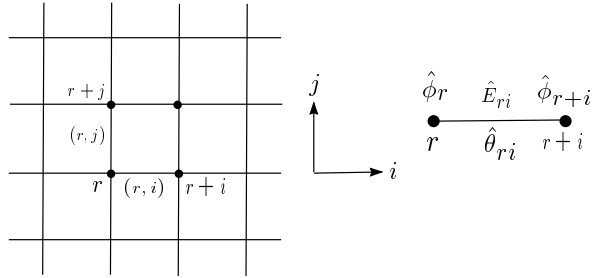}\\
  \end{center}
\caption{
$i-j$ plane of the 3d lattice and $\hat{\phi}_r$ defined on the site $r$
and $\hat{\theta}_{ri}$ and $\hat{E}_{ri}$ defined on the link $(ri)=(r,r+i)$.}
\label{lattice}
\end{figure}

Let us start with an example of Hamiltonian system defined on a 
three-dimensional ($d=3$) spatial cubic lattice with lattice spacing $a$. 
Its sites are labeled by $r$, and 
its links (bonds) are labeled as $(r,i)$ where $i$
is the direction index $i=1,2,3$.
We sometimes use $i$ also for the unit vector in the $i$th-direction,
such as $(r,i)=(r,r+i)$ (See Fig.\ref{lattice}).

One may put matter fields on each  site $r$, which 
may be bosons/fermions composing materials in question, e.g., electrons,
atoms in an optical lattice, Coope-pair field of SC,
spins, or even more complicated ones.
For definiteness, we put below a one-component canonical nonrelativistic 
boson field\cite{boson}, whose annihilation and creation operators $\hat{\phi}_r,
\hat{\phi}_r^\dag$ satisfy
\be
[\hat{\phi}_r, \hat{\phi}^\dag_{r'}]=\delta_{rr'}.
\ee 
Gauge fields are put on each link $(r,i)$  in LGT (See Fig.\ref{lattice}). 
They may be associated with 
the gauge group U(1), SU(N), etc, and    
meditate interactions between matters. They are viewed as connections 
in differential geometry that relate two local frames of the 
internal space of matter field at neighboring lattice sites $r$ and $r+i$. 
For definiteness, we consider below a U(1) gauge field operator
$\hat{\theta}_{ri}\ (= \hat{\theta}^\dag_{ri})$.
This is viewed  as the lattice version of well-known vector potential field
$\hat{A}_i(r)$ (we use the same letter $r$ for the coordinate of the continuum space)
describing the electromagnetic(EM) interaction, but may have different origins,
e.g., mediating interactions between holons and spinons,
as explained in Section 1.
The  momentum operator $\hat{E}_{ri}$ conjugate to $\hat{\theta}_{ri}$
describes the electric field and satisfy
\be
[\hat{\theta}_{ri}, \hat{E}_{r'j}]=i\delta_{rr'}\delta_{ij},\
[\hat{\theta}_{ri}, \hat{\theta}_{r'j}]=[\hat{E}_{ri}, \hat{E}_{r'j}]=0.
\label{eri}
\ee

To understand the LGT Hamiltonian, it is helpful to start with
a model Hamiltonian $\hat{H}_{\rm c}$ defined in the continuum space. 
Let us consider the following Ginzburg-Landau(GL)-type model:
\be
H_{\rm c}&=&
\int d^3r\Big[\sum_i
|D_i\hat{\phi}(r)|^2 +m^2\hat{\rho}(r)+\lambda\hat{\rho}(r)^2
+\frac{1}{2}\sum_i\hat{E}_i(r)^2+\frac{1}{4}\sum_{i,j}\hat{F}_{ij}(r)^2\Big],\nn
D_i&=&\partial_i+ig \hat{A}_i(r),\ \hat{\rho}(r) = 
\hat{\phi}^\dag(r)\hat{\phi}(r),\ 
\hat{F}_{ij}(r)=\partial_i\hat{A}_j(r)-\partial_j\hat{A}_i(r),
\label{Hc}
\ee
where $D_i$ is the covariant derivative expressing the minimal coupling
of $\phi$-particle with the EM field, $g$ is the coupling constant (charge
of $\phi$-particle\cite{charge}), 
$m, \ \lambda$ are GL parameters whose meaning is explained later, 
and $\hat{F}_{ij}(r)$ is the magnetic field $B_i(r)=\epsilon_{ijk}F_{jk}(r)$. 
One may check that the each term  of  $H_{\rm c}$  
is invariant under the local gauge transformation,
\be
\hat{\phi}(r) \rightarrow e^{i\Lambda(r)} \hat{\phi}(r),\  
\hat{A}_i(r)\rightarrow \hat{A}_i(r)-\frac{1}{g}\partial_i \Lambda(r),\
\hat{E}_i(r)\rightarrow \hat{E}_i(r),
\label{localgt}
\ee
for an arbitrary function $\Lambda(r)$.
By using the canonical commutation relations 
$[\hat{\phi}(r),\hat{\phi}^\dagger(r')]=
\delta^{(3)}(r-r')$ and $[\hat{A}_i(r), \hat{E}_j(r')]=i\delta_{ij}\delta^{(3)}(r-r'),$
one may check that its generator $\hat{Q}(r)$ is given by
\be
\hat{Q}(r) = \sum_i \partial_i \hat{E}_i(r) -g\hat{\rho}(r),
\ee
and $\hat{Q}(r)$ commutes with $\hat{H}_{\rm c}$,
$[\hat{Q}(r), \hat{H}_{\rm c}] = 0$.
So one may define the physical states $|\rm phys\ra$
so that they are eigenstates of $\hat{Q}(r)$
(superselection rule). In the usual case,
i.e., when there are no  external electric field, one imposes the 
local constraint,
\be
\hat{Q}(r)|\rm phys\ra =0,
\ee
which leads to the lattice version of the Gauss' law $\hat{Q}_r=0$
[See Eq.(\ref{physical}].

Now we consider the Hamiltonian on the cubic lattice.
In LGT, the operator $\hat{\theta}_{ri}$ defined on link $(r,i)$ 
is assumed to approach as
$\hat{\theta}_{ri}\to a g \hat{A}_i(r)$ as the lattice spacing $a$ becomes small.
Furthermore, as we shall see, the natural U(1) gauge field operator in LGT
is not $\hat{\theta}_{ri}$ but its exponential,    
\be
\hat{U}_{ri}\equiv \exp(i\hat{\theta}_{ri}),\ \hat{U}_{ri}^\dag\hat{U}_{ri}=\hat{1},
\label{UA}
\ee
and so $\hat{\theta}_{ri}$ is viewed as an angle operator (we discuss this point 
in detail later). 
Then, by using Eq.(\ref{eri}) we have 
$\hat{E}_{ri}\to a^2 E_i(r)$ as 
$a\to 0$ (note $\hat{E}_{ri}$ is dimensionless and $\delta^{(3)}(r)\simeq
\delta_{r0}a^{-3}$) and 
\be
[\hat{E}_{ri}, \hat{U}_{r'j}]=\delta_{ij}\delta_{rr'}\hat{U}_{ri}.
\label{EU}
\ee
(Note that Eq.(\ref{eri}) implies the replacement $\hat{E}_{ri}\to
-i\partial/\partial \theta_{ri}$.)
To construct a set of eigenstates satisfying 
the completeness conditions, let us start with
a pair of operators $\hat{U}$ and $\hat{E}$ sitting on certain link
(we suppress their link index). 
Then we have
\be
&&\hat{U}|\theta\ra=e^{i\theta}|\theta\ra,\ \theta \in [0,2\pi\},\
\hat{E}|n\ra=n|n\ra,\ n \in {\bf Z},\nn
&& \la \theta|n\ra= \frac{e^{in\theta}}{\sqrt{\mathstrut{2\pi}}},\ 
\hat{1}_U =\int_0^{2\pi} d\theta |\theta\ra\la \theta|,\ 
\hat{1}_E=\sum_{n= -\infty}^{\infty}|n\ra\la n|.
\label{uande}
\ee
So $\hat{\theta}$ and $\hat{E}$ correspond to the position and momentum
operators in the ordinary quantum mechanics respectively, but the significant 
difference is that the theory is taken to be {\it compact}, i.e.,  
the physical state $|\rm phys\ra$ is  imposed to be periodic in $\theta$,
\be
\la\theta |\rm phys\ra = \la\theta+2\pi |\rm phys\ra,
\label{periodicity}
\ee
which implies that the eigenvalues are an angle $\theta \in [0,2\pi)$ 
defined by mod $2\pi$, and integers $n$, instead of two real numbers.
Because one may write 
\be
\hspace{-0.7cm}
|n\ra = \hat{1}_U |n\ra 
=\int  \frac{d\theta}{\sqrt{\mathstrut{2\pi}}}e^{in\theta}|\theta\ra
= e^{in\hat{\theta}}\cdot\int  \frac{d\theta}{\sqrt{\mathstrut{2\pi}}}|\theta\ra
=  e^{in\hat{\theta}} |n=0\ra =\hat{U}^n |n=0\ra,
\ee
$\hat{U}^{(\dag)}$ is just the creation (annihilation) operator of the 
electric field of unit electric flux (the signature of $n$ distinguishes 
the direction of the flux), which is called a ``string bit". 
These relations remind us the analogy with the angular momentum
operators $\vec{L}$ as $\hat{E} \leftrightarrow \hat{L}_z 
(n \leftrightarrow L_z/\hbar), \hat{U}
\leftrightarrow \hat{L}_x +i\hat{L}_y, \theta \leftrightarrow \varphi$
(azimuthal angle) with the limit of large angular momentum $\ell\to \infty$.
We note that there holds
the uncertainty principle,
\be
\Delta \theta\cdot \Delta n \geq \frac{1}{2}.
\label{uncertainity}
\ee

The Hamiltonian $\hat{H}$ of LGT should  (i) reduce to $\hat{H}_c$ of 
Eq.(\ref{Hc}) 
in the naive continuum limit $a\to 0$, and (ii) respect the U(1) local gauge 
symmetry as in the continuum.
A simple example of $\hat{H}$ satisfying these two conditions is given by
\be
\hat{H}&=&t\sum_{r,i}
(\hat{\phi}^\dagger_{r+i}-\hat{U}_{ri}\hat{\phi}^\dagger_r)
(\hat{\phi}_{r+i}-\hat{U}^\dag_{ri}\hat{\phi}_r)
+\sum_r V(\hat{\phi}_r)+\hat{H}_{\rm g},\ 
\nn
\hat{H}_{\rm g}&=&\frac{g^2}{2a}\sum_{r,i}(\hat{E}_{ri})^2+\frac{1}{2g^2a}\sum_{r}\sum_{i<j}\Big(\hat{U}_{rj}
\hat{U}_{r+j,i}
\hat{U}^\dag_{r+i,j}\hat{U}^\dag_{ri}+{\rm H.c.}-2\Big).
\label{Hgauge}
\ee
Here $V(\hat{\phi}_r)$ represents the self-interaction of $\hat{\phi}_r$.
The naive continuum limit of $\hat{H}$ is checked as
$\hat{H} \to \hat{H}_{\rm c} +O(a^2)$ with a suitable choice of $V(\hat{\phi}_r)$ 
by using the following relations as $a\to 0$:
\be
&&\hat{\phi}_r \to a^{\frac{3}{2}} \hat{\phi}(r),\ 
\hat{U}_{ri} \to e^{iga \hat{A}_i(r)}\simeq 1+ 
iga \hat{A}_i(r) -\frac{a^2g^2}{2}
(\hat{A}_i(r))^2,\nn
&&\hat{E}_{ri}\to \frac{a^2}{g}\hat{E}_i(r),\
\sum_r \to \frac{1}{a^3}\int d^3r.
\ee
In particular, the last $UUUU$-term reduces to the magnetic term $FF$
as
\be
\hspace{-1cm}
&&UUUU+{\rm c.c.}\to\exp[iga(A+A-A-A)]+{\rm c.c.}\nn
\hspace{-1cm}&=&\exp[iga^2(\partial_iA_j(r)-\partial_jA_i(r))]+{\rm c.c.}
\to 2\cos(ga^2F_{ij}(r))\to 2-g^2a^4F^2_{ij}(r).
\label{magnetic}
\ee 

For gauge invariance, one may define the gauge transformation on the lattice as
\be
\hat{U}_{ri}\to V^\ast_{r+i}\hat{U}_{ri}V_r,\ V_r = e^{i\Lambda_r} 
\in {\bf U(1)},\ \hat{\phi}_r \to V_r \hat{\phi}_r,
\label{gaugetr}
\ee
which reduces to Eq.(\ref{localgt}) by scaling $\Lambda_r\to \Lambda(r)$ as $a\to 0$,
and check that each term of $\hat{H}$ is invariant
under Eq.(\ref{gaugetr}).
We define the physical states in the same manner as in the continuum case,
\be 
Q_r\equiv\sum_i \nabla_i\hat{E}_{ri}-\hat{\phi}^\dagger_r\hat{\phi}_r,\ 
[\hat{H}, \hat{Q}_r]=0,\ \hat{Q}_r|\rm phys\ra =0,
\label{physical}
\ee
where the forward difference operator $\nabla_i$ is defined as
$\nabla_if_r\equiv f_{r+i}-f_r$. Both $\nabla_i\hat{E}_{ri}$ 
and $\hat{\phi}^\dagger_r\hat{\phi}_r$ have integer eigenvalues.
We note that $\hat{H}$ of Eq.(\ref{Hgauge}) has the periodicity $\hat{\theta}_{ri}\to
\hat{\theta}_{ri}+2\pi$, which may be regarded as a
gauge symmetry under a special gauge transformation
 $\Lambda_r=2\pi r_i$.

\subsection{Physical properties of LGT}

Let us discuss the expected properties of the model (\ref{Hgauge}). 
First, we focus on the case of pure gauge theory $\hat{H}_{\rm g}$
without matter field. In te continuum theory,
the corresponding system, 
\be
\hat{H}_{\rm g}\to \frac{1}{2}\int d^3r \sum_i\Big(
\hat{E}_i(r)\hat{E}_i(r)+\hat{B}_i(r)\hat{B}_i(r)\Big),
\ee
is well known to describe the ensemble of free photons. In contrast, 
$\hat{H}_{\rm g}$ on the lattice is an {\it interacting} theory.
One can confirm it  from $\cos(ga^2F_{ij}(r))$ in Eq.(\ref{magnetic}). 
It contains, in addition to the leading $F_{ij}(r)^2$ term,
$F_{ij}(r)^4$  and higher interaction terms describing scattering of gauge bosons.
These terms are 
traced back to the compactness 
of the model (the periodicity (\ref{periodicity})).

For large $g \gg 1$, which is called the strong coupling region,
one may take the electric term as the unperturbed
Hamiltonian and the magnetic term as a perturbation. 
By recalling Eq.(\ref{physical}), the unperturbed eigenstate $|\rm eigen\ra$ is
the eigenstate of $\hat{E}_{ri}$ with their eigenvalues satisfying
the divergenceless condition. A simple example is an electric flux
of strength $n$ along a closed loop C on the 3D lattice.
The general eigenstate may be composed of such loops. 
So one may write 
\be
|{\rm eigen}\ra = \prod_k \prod_{\ell_k \in C_k}(\hat{U}_{\ell_k})^{n_k}|0\ra,
\ee
where $C_k$ denotes the $k$-th flux loop, $\ell_k$ denotes links composing $C_k$,
and $n_k$ is the strength of the $k$-th flux.
If there are two external sources of charge $g$ at $r_1$ and $-g$ at $r_2$,
the lowest-energy state is an electrix flux state forming a straight line
$\tilde{C}_{12}$ (on the lattice) connecting these two charges 
and written as
\be
\prod_{\ell\in \tilde{C}_{12} } \hat{U}_{\ell}|0\ra,
\label{pair}
\ee
(See Fig.{\ref{eflux}a).
This state has an energy $g^2/(2a)\times $ number of links in $C_{12}$, thus
proportional to the distance $r$. If these two charges are quark and antiquark,
isolation of each quark from the  other implies $r\to\infty$ and costs
infinite amount of energy, being impossible to be realized.
This is  Wilson's explanation of quark confinement\cite{wilson}.

\begin{figure}[b]
  \begin{center}
      \includegraphics[width=8cm]{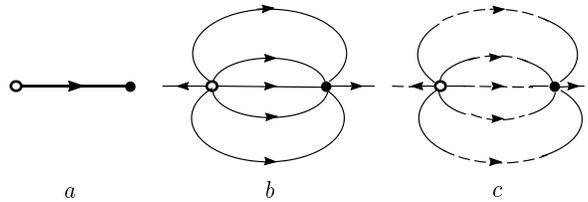}\\
  \end{center}
\caption{
Illustration of electric flux lines between two oppositely charged particles
in the three phases.
(a) confinement phase; (b) Coulomb phase; (c) Higgs phase.
Dashed lines in (c) imply that the strength of fluxes are not conserved
as Eq.(\ref{physical}) with $\la\hat{\phi}^\dag_r\hat{\phi}_r\ra = {\rm const}\
 \neq 0$ shows.}
\label{eflux}
\end{figure}

On the other hand, for the weak-coupling region, $g\ll 1$, the magnetic term
is leading. Up to the gauge transformation, 
the energy eigenstates in the limit $g \to 0$ should be the eigenstate of 
$\hat{\theta}_{ri}$ in contrast to the strong-coupling limit.
By the uncertainty principle (\ref{uncertainity}),
the eigenstate there should be a superposition of electric fluxes
of various strength and locations as in the ordinary classical 
Coulombic electromagnetism
shown in Table.1. So the physical state corresponding to the state 
considered in Eq.(\ref{pair}) is replaced by
\be
\sum_{C_{12}}
\Gamma(C_{12})\prod_{\ell\in C_{12}} \hat{U}_{\ell}
|0\rangle,
\label{pair2}
\ee
where the coefficient $\Gamma(C_{12})$ is the weight for the path $C_{12}$
connecting the sites  1 and 2. For the lowest-energy state, 
$\Gamma(C_{12})$ is to be determined by the minimum-energy condition,
and the energy stored in the  two charges  here should be proportional to $1/r$ 
as in the Coulomb potential energy (See Fig.\ref{eflux}b).

The above qualitative consideration may indicate that there exists a phase boundary
separating the strong and weak-coupling regions.
Dynamics of the pure gauge system $H_g$ in (\ref{Hgauge}) is quite nontrivial 
and careful investigation such as numerical study is required to obtain the
phase diagram.
At present, it is known that the 3D pure U(1) gauge system $H_g$ 
has  two phases;
The phase at strong coupling region is called the confinement phase, and
the phase at weak coupling region is called Coulomb phase.
The phase transition is said to be weak first order\cite{3+1d}.
These two phases are characterized by the behavior of the potential energy
$V(r)$ for a pair of static charges of opposite signs and 
separated by a distance $r$; $\propto r$ or $\propto 1/r$.
Also they are distinguished by the fluctuation of gauge field $\hat{\theta}_{ri}$,
$\Delta \theta$, as $\Delta \theta \gg 1$ in the confinement phase
and $\Delta \theta \ll 1 $ in Coulomb phase.  
We note that, for the 2D spatial lattice in contrast, there are no phase 
transitions, and only the confinement phase exists\cite{polyakov}.

Now let us include the matter part and consider the full 
Hamiltonian $\hat{H}$ of Eq.(\ref{Hgauge}).
By referring to Eq.(\ref{Hc}), one may write $V(\hat{\phi}_r)$ as 
\be
V(\hat{\phi}_r)= \lambda (\hat{\phi}^\dag_r\hat{\phi}_r- v)^2.
\label{phi4}
\ee
Let us recall that the GL theory with
the  gauge interaction being turned off by setting $g=0$
and for sufficiently large and positive $\lambda$,
the system exhibits a second-order phase transition by varying $v$. 
It is signaled by the order parameter $\phi = \langle \hat{\phi}_r\rangle$, 
(or more strictly defined as 
$|\phi|^2 \equiv \lim_{|r-r'|\to\infty}\la \hat{\phi}^\dag_r 
\hat{\phi}_{r'}\ra$),
and separates the ordered (condensed) phase 
$\langle\phi\rangle\neq 0$ and the disordered
phase $\langle\phi\rangle=0$. In the mean-field theory,
the transition temperature $T_{\rm c}$ is determined 
as the vanishing point of $v$ as $v\propto (T_{\rm c}-T)$. 
However, the fluctuations of the phase degree of freedom $\varphi_r$ of
$\hat{\phi}_r = |\hat{\phi}_r| \exp(i\varphi_r)$, which are controlled 
by the hopping $t$-term of Eq.(\ref{Hgauge}), 
play an essentially important role for realization of a condensation.
In fact, these two phases may be characterized by the magnitude of fluctuation
$\Delta\varphi$ of $\varphi_r$ as $\Delta\varphi \ll 1$ (ordered phase)
and $\Delta\varphi \gg 1$(disordered phase).
To obtain the correct transition temperature $T_{\rm c}$,
more detailed study such as MC simulation is required as we discuss later.

For the full fledged gauge-matter system $\hat{H}$ of Eq.(\ref{Hgauge}), 
one may naively consider four combinations
such as ($\Delta\theta \ll 1\ {\rm or} \gg 1$) and 
($\Delta\varphi \ll 1\ {\rm or} \gg 1$) for a possible phase. 
However, one may easily recognize that the combination
$\Delta\varphi \ll 1$ and $\Delta\theta \gg 1$ is impossible.
This is because $\hat{\varphi}_r$ appears in the hopping term with the combination
$\nabla_i\hat{\varphi}_r -\hat{\theta}_{ri}$ and the condition $\Delta\theta \gg 1$
destroys the phase coherence $\nabla_i\hat{\varphi}_r \simeq 0$.
In fact, the numerical study and the mean field theory predict
the three phases listed in Table 1; 
(1) confinement phase $\Delta\theta \gg 1, \Delta\varphi \gg 1$,
(2) Coulomb phase $\Delta\theta \ll 1, \Delta\varphi \gg 1$,
(3) Higgs phase $\Delta\theta \ll 1, \Delta\varphi \ll 1$.

Let us see this point in some details because it helps us to understand 
the property of each phase appeared in an effective  LGT of various condensed-matter 
systems.  

\begin{enumerate}
\item The confinement phase appears for the parameter region of
small $t$ [$t$ is the hopping amplitude in Eq.(\ref{Hgauge})] and large $g$  regardless of the values of $\lambda$ and $v$.
In a sense, this phase corresponds to the high-temperature ($T$) region
and there exist no long-range orders.
The parameter region of the confinement phase is enlarged by decreasing
a value of $v$.
For pure U(1) gauge theory without matter field $\hat{\phi}_r$,
a potential $V(r)$ for a pair of oppositely charged static sources
separated with distance $r$ behaves as $V(r)\propto r$ as explained.

\item The Coulomb phase appears for small $g$ and small $t$.
Fluctuations of the gauge field is suppressed by the 
plaquette terms in Eq.(\ref{Hgauge}) and 
the compactness of $\hat{\theta}_{ri}$ plays no role.
The potential of a pair of two static charges is given by the Coulomb
law, which is $V(r)\propto 1/r$ for 3D space.
\item In the Higgs phase, a coherent condensation of the boson field
$\hat{\phi}_r$ takes place. 
Then the hopping term in Eq.(\ref{Hgauge}) gives the following term
of $U_{x\mu}$,
\be
t\hat{\phi}^\dag_{r+i}\hat{U}^\dag_{ri}\hat{\phi}_r+{\rm H.c.}
\to tv\ (\hat{U}_{ri}^\dag+U_{ri})+\cdots \simeq -t v a^2 g^2
\hat{\theta}^2_{ri}+\cdots.
\ee
As a result, the gauge field acquires a mass $m_\theta=\sqrt{\mathstrut{2tv}}$ 
and a force mediated by the 
gauge field $\hat{\theta}_{ri}$ becomes short range like 
$\langle \hat{\theta}_{r+R i,j}\hat{\theta}_{tj}\rangle=e^{-m_\theta R}$
 (See Fig.\ref{eflux}c).
The Higgs phase is realized for large value of $t$ and a finite $v$
(in addition to small $g$).
Superconducting (SC) phase is a kind of the Higgs phase in which 
the boson field $\hat{\phi}_r$ describes Cooper pairs, and 
the gauge field becomes short-ranged by the Meissner effect.
\end{enumerate}

So far, we studied the compact LGT. In some literatures, so called
noncompact LGT is considered\cite{cvsnc}. The noncompact U(1) LGT is
given by the following Hamiltonian for the gauge part,
\be
\hat{H}'_{\rm g}&=&\frac{g^2}{2a}\sum_{r,i}(\hat{E}_{ri})^2
+\frac{1}{2g^2a}\sum_{r}\sum_{i<j}(\nabla_i\hat{\theta}_{rj}-
\nabla_j\hat{\theta}_{ri})^2.
\label{Hgaugenc}
\ee
The eigenvalue of $\hat{\theta}_{ri}$ now runs in $(-\infty,\infty)$
and $\hat{H}'_{\rm g}$ loses the periodicity and describes a set of free photons. 
For small fluctuations $\Delta\theta \ll 1$,
the compact pure gauge theory (\ref{Hgauge}) reduces to 
above $\hat{H}'_{\rm g}$.
However, for large fluctuations, these two models behave quite differently
and they generally have different phase diagrams.
In the noncompact gauge theories with the gauge part (\ref{Hgaugenc}),
the Coulomb and Higgs phases are realized but the confinement phase is
impossible because of the suppression of large fluctuations.
In other words, the large fluctuations of $\hat{\theta}_{ri}$ 
in the confinement phase are achieved by large-field configurations 
called topological excitations, the typical examples of which are  
instantons and monopoles. They are allowed only for a system
with periodicity such as compact LGT.
To describe a SC phase transition with the ordinary EM
interactions, one should use the noncompact U(1) gauge theory.
The gauge field $\hat{\theta}_{ri}$ there is nothing but the vector potential 
for the electro-magnetism, and the single-component boson field 
$\hat{\phi}_r$ corresponds to the $s$-wave Cooper-pair field as mentioned.
Therefore, as we have explained above, the Higgs phase is nothing but the SC phase.
In order to describe a multi-component SC state such as  the $p$-wave SC, 
introduction of a multi-component boson field is necessary.
This system will be discussed rather in detail in the subsequent sections.

\subsection{Partition function in the path-integral representation 
on the Euclidean lattice}

Let us consider the partition function $Z$ for $\hat{H}$ of Eq.(\ref{Hgauge}) 
at the temperature $T$,
\be
Z= {\rm Tr} \hat{P}\exp(-\beta \hat{H}), \ \beta \equiv \frac{1}{k_B T}, \ 
\hat{P}\equiv\prod_r\delta_{\hat{Q}_r,0},
\ee
where Tr is over the space of all the values of $\hat{Q}_r$,
and $\hat{P}$ is the projection operator to the physical states.
As usual, we start by factorizing the Boltzmann factor
into $N$ factors but with care of $\hat{P}$  as 
\be
\hat{P}\exp(-\beta \hat{H})=[\hat{P}\exp(-\Delta\beta \hat{H})]^N,\
\Delta \beta \equiv \frac{\beta}{N},
\ee
where we used $\hat{P}^2=\hat{P}, [\hat{P}, \hat{H}]=0$, and then
insert the complete set between every successive factors.

For $\hat{\phi}_r$, we use the following coherent states $|\phi\ra$:
\be
&&\hat{\phi}|\phi\ra=\phi|\phi\ra,
|\phi\ra = \exp(-\bar{\phi}\hat{\phi}+\hat{\phi}^\dag\phi)|0\ra,\ 
\hat{\phi}|0\ra=0,\ 
\la \phi'|\phi\ra=\exp(-\frac{\bar{\phi}'\phi'}{2}-\frac{\bar{\phi}\phi}{2}
+\bar{\phi}'\phi'),\nn
&& |\{\phi\}\ra=\prod_r|\phi_r\ra,\ \int d^2\phi =\prod_r \int \frac{d^2\phi_r}{\pi},\
\hat{1}_\phi =\int d^2\phi |\{\phi\}\ra\la \{\phi\}|,
\ee
where $\phi$ is a complex number 
and $\bar{\phi}$ is its complex conjugate 
(The bar denotes the complex-conjugate quantity).

The eigenstates and the completeness of gauge field for 
the entire lattice are written by using Eqs.(\ref{uande}) as
\be
&&\hat{1}_\theta =\int d\theta  |\{\theta\}\ra\la \{\theta\}|,\
\hat{1}_n= 
\sum_n |\{n\}\ra\la \{n\}|,\ 
|\{\theta\}\ra = \prod_{r,i} |\theta_{ri}\ra,\ |\{n\}\ra = 
\prod_{r,i} |n_{ri}\ra, \nn
&&\int d\theta =\prod_{r,i} \int d\theta_{ri},\
\sum_n =\prod_{r,i} \sum_{n_{ri} \in {\bf Z}}.
\ee
(We abbreviate the symbol $\otimes$ of tensor product in $|\{\theta\}\ra$
and $|\{n\}\ra$.)

For example, the matrix element of $\hat{P}$ is calculated as
\be
&&\delta_{\hat{Q}_r,0}=\int_0^{2\pi} \frac{d\theta_{r0}}{2\pi} 
\exp(i\theta_{r0}\hat{Q}_r),\
\la\{\phi', n'\}|\hat{P}| \{\phi, n\}\ra =\prod_r
\int \frac{d\theta_{r0}}{2\pi}\prod_i\delta_{n'_{ri}n_{ri}}\cdot \nn
&&
\prod_r\exp(-\frac{1}{2}\bar{\phi}'_r\phi'_r-\frac{1}{2}
\bar{\phi}_r\phi_r+\bar{\phi}'_re^{-i\theta_{r0}}\phi_r) 
\exp(i\theta_{r0}\sum_i\nabla_iE_{ri}).
\ee
The Lagrange multiplier 
$\theta_{r0}$ will be interpreted as the imaginary-time component of gauge field.
Then we obtain the following expression of $Z$ for sufficiently large $N$:
\be
\hspace{-0.2cm}
&&Z =\int [dU] \int [d\phi] \sum_{\{n\}} \exp(A_H),\ 
[dU]=\prod_{x,\mu}dU_{x\mu},\ [d\phi]=\prod_x d\phi_x,
\sum_{\{n\}} =\prod_{x,i}\sum_{n_{xi}},  
\nn 
\hspace{-0.2cm}
&&A_H =  \Delta \beta\sum_x \Big[\frac{i}{\Delta \beta}
\sum_i (E_{xi}\nabla_0\theta_{xi}+
\theta_{x0}\nabla_iE_{xi})
-\frac{1}{ \Delta \beta}\bar{\phi}_{x+0}(\phi_{x+0}-e^{-i\theta_{x0}}\phi_x)
-V(\phi_x)\nn
\hspace{-0.2cm}
&&-t\sum_i
|\phi_{x+i}-\bar{U}_{xi}\phi_x|^2-
\frac{g^2}{2a}\sum_i E_{xi}^2 +\frac{1}{2g^2 a}\sum_{i<j}
(U_{xj}U_{x+j,i}\bar{U}_{x+i,j}\bar{U}_{xi}+{\rm c.c.})\Big].
\label{zah}
\ee
where we assume that $V(\hat{\phi}_x)$ is normal ordered.
In Eq.(\ref{zah}) we introduced the imaginary-time coordinate
$x_0 (= 0, \cdots, N-1)$, the site index of 4D Euclidean lattice 
$x \equiv (x_0,r)$ with $r=(x_1, x_2, x_3)$, and
its direction index/unit vector $\mu =0,1,2,3$. 
In path integral, every field $\phi_x,\theta_{x\mu},n_{xi}$ 
is defined on each time slice $x_0$ together with the spatial site $r$.
We note that 
$\phi_x,\theta_{x\mu}$ are periodic under $x_0 \to x_0+N$.
The 4-component U(1) gauge field $U_{x\mu} =\exp(i\theta_{x\mu})$ and
its complex conjugate $\bar{U}_{x\mu}$ are sitting 
on the link $(x,x+\mu)$.

Next, we carry out the summation over $n_{xi}$ to obtain\cite{uubar}
\be
Z &=& \int [dU][d\phi]\  \exp(A),\ 
A =  \sum_x \Big[
-\bar{\phi}_{x+0}(\phi_{x+0}-\bar{U}_{x0}\phi_x) -\Delta\beta V(\phi_x)\nn
&&-t\Delta\beta\sum_i
|\phi_{x+i}-\bar{U}_{xi}\phi_x|^2 +\frac{1}{2}
\sum_{\mu<\nu}c_{2\mu\nu}(U_{x\nu}U_{x+\nu,\mu}\bar{U}_{x+\mu,\nu}\bar{U}_{x\mu}+{\rm c.c.})\Big],\nn
c_{2ij}&=&\frac{\Delta\beta}{g^2a},\ c_{20i}=\frac{a}{g^2\Delta\beta},\
 U_{x0}\equiv\exp(i\theta_{x0}).
\label{nrlgt}
\ee
Here we made the following replacement in $\sum_{n_{xi}}$ 
respecting the periodicity in $\theta$:
\be
\hspace{-0.5cm}\sum_n e^{-a n^2+in\theta} \simeq  \int_{-\infty}^\infty dn\
e^{-a n^2+in\theta} = C\exp(-\frac{\theta^2}{4a})
\to C'\exp(\frac{1}{2a}\cos\theta).
\ee
The action $A$ and the measure $[dU][d\phi]$ in Eq.(\ref{nrlgt} are invaraiant
under the U(1) gauge transformation rotating the local internal coordinate
of complex field at each site $x$ on the 4-dimensional lattice,
\be
U_{x\mu}\to \bar{V}_{x+\mu}U_{x\mu}V_x,\ V_x = e^{i\Lambda_x} 
\in {\bf U(1)},\ \phi_x \to V_x \phi_x,
\label{gaugetr2}.
\ee

\subsection{Phase structure and Correlation functions}

Phase structure of various gauge-field models has been investigated
by both analytic and numerical methods.
In particular, for the gauge-Higgs model, which describes the SC 
phase transitions, interesting phase diagrams have been obtained.
To this end, the numerical study by the Monte-Carlo (MC) simulations
plays a very important role as they provide us reliable results
including all nonperturbative effects.
Here we present some known schematic phase diagrams restricting the system to 
the {\it relativistic Higgs coupling} and the frozen radial degrees of freedom 
of $\phi_x$ field, i.e., $\phi_x=\exp(i\varphi_x)$,
but with spatial dimensions $d=2$ and 3 (dimension of the 
corresponding Euclidean lattice $D=d+1$ is 3 and 4).
Explicitly, we consider the following partition function;
\be
\hspace{-0.9cm}
Z&=&\int[d\varphi][d\theta]_G\exp(A),\ 
A = c_1\sum_{x,\mu}\cos(\varphi_{x+\mu}+\theta_{x\mu}-\varphi_x)+ A^{(')}_g,\nn
\hspace{-0.9cm}
A_g&=& c_2\!\!\!\sum_{x,\mu < \nu}\!\!\cos\theta_{x\mu\nu},\
A'_g= -\frac{c_2}{2}\!\!\sum_{x,\mu < \nu}\!\theta^2_{x\mu\nu},\
\theta_{x\mu\nu}\!\equiv\!\theta_{x\nu}\!+\theta_{x+\nu,\mu}\!-
\theta_{x+\mu,\nu}\!-\theta_{x\mu},
\label{u1lgt}
\ee
where $\mu=0,\cdots, d,$ and $c_1$ and $c_2$ are the parameters. 
$A_g$ is for the compact LGT and $A'_g$ is for the noncompact LGT.
Each term of the action $A$ is depicted in Fig.\ref{energy}.
This model is sometimes called the Abelian Higgs model\cite{savit}.
Note that the Higgs coupling (the $c_1$ term) along the imaginary time
$\mu=0$ direction has the same form as the spatial directions,
i.e., every direction has couplings both in the positive and negative directions. 
This is in contrast to the nonrelativistic model (\ref{nrlgt})
in which the coupling for $\mu=0$ is only in the positive direction.
In short, the coupling in the negative $\mu=0$ direction describes
the existence of antiparticles which are allowed in 
the relativistic theory but not in nonrelativistic theory\cite{anti}.

\begin{figure}[t]
  \begin{center}
      \includegraphics[width=8cm]{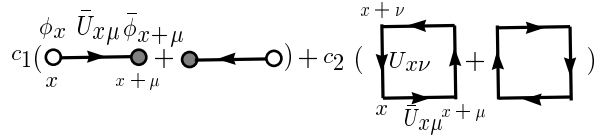}\\
  \end{center}
\caption{
Illustration of the action $A$ of the compact LGT (\ref{u1lgt}).
Open (filled) circle denotes $\phi_x=\exp(i\varphi_x) (\bar{\phi}_x)$ and 
straight line with an arrow in the negative (positive) $\mu-$direction denotes 
$U_{x\mu}=\exp(i\theta_{x\mu}) (\bar{U}_{x\mu})$}
\label{energy}
\end{figure}

Here we comment on the integration measure $[d\theta]_G$ and gauge fixing.
The path integral (\ref{u1lgt}) involves the same contributions
from a set of configurations of $\{\varphi,\theta\}$ (so called gauge copies)
that are connected with each other by gauge transformations. So
one may reduce this redundancy by choosing only one representative from each 
set of gauge copies without changing physical contents. 
This procedure is called gauge fixing and achieved by inserting
the gauge fixing function $G(\{\varphi,\theta\})$, which is not gauge invariant
under Eq.(\ref{gaugetr2}),
to the measure,
$[d\varphi][d\theta]\to [d\varphi][d\theta]_G\equiv 
G(\{\varphi,\theta\})[d\varphi][d\theta]$.
The difference between the gauge fixed case $[d\theta]_G$ and 
the nonfixed case $[d\theta]$  
appears as a multiplicative factor in $Z$, 
\be
Z_{\rm nonfixed} = C_{\rm gv}Z_{\rm fixed},\ C_{\rm gv} = 
\prod_{x,\mu}\int d\Lambda_{x\mu},
\label{gv}
\ee
where the constant $C_{\rm gv}$ is called gauge volume.   
In the compact case, the  region of $\theta_{x\mu}$ and 
$\Lambda_{x\mu}$ are both compact $[0,2\pi)$. Therefore
$C_{\rm gv}$ is finite and the gauge fixing is irrelevant
(either choice of fixing gauge or not will do).
In the noncompact case, the region of
$\theta_{x\mu}$ and $\Lambda_{x\mu}$ is $(-\infty, \infty)$,
so $C_{\rm gv}$ diverges. Thus, one must fix the gauge in formal 
argument in the noncompact case,
$[d\varphi][d\theta]_G$. 
We said ``in formal argument" here because the choice $G(\{\varphi,\theta\})=1$ 
works even in the noncompact case as long as one 
calculate the average of gauge-invariant quantities, e.g.,
in the MC simulations.
This is because the average does not suffer from the overall factor in $Z$. 
We shall present in Sect.4.1. another reason supporting this point, 
which is special in MC simulations. 

\begin{figure}[h]
 \begin{minipage}{0.49\hsize}
  \begin{center}
      \includegraphics[width=3.5cm]{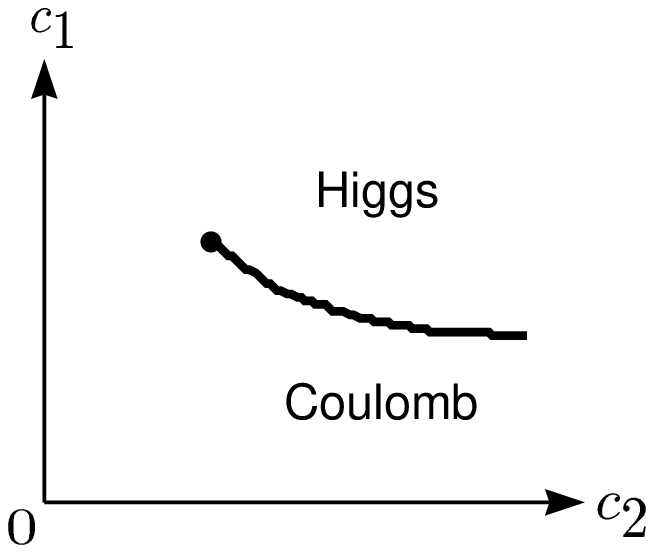}\\
      (a) Noncompact 3D
  \end{center}
 \end{minipage}
 \begin{minipage}{0.49\hsize}
  \begin{center}
      \includegraphics[width=3.5cm]{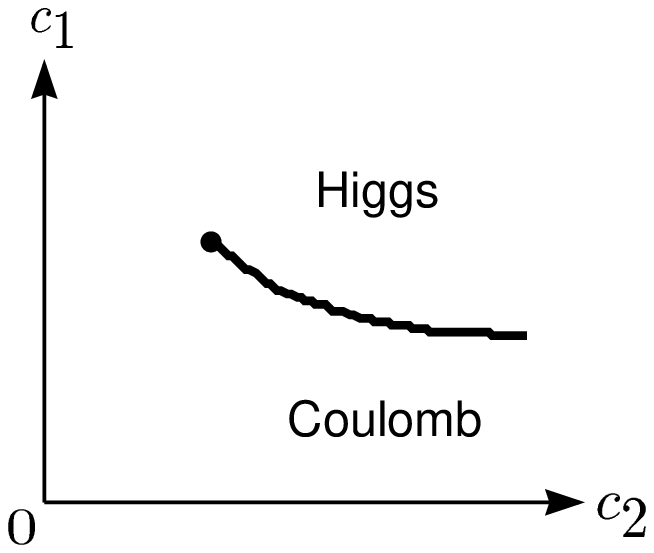}\\
      (b) Noncompact 4D
  \end{center}
 \end{minipage}\\
 \begin{minipage}{0.49\hsize}
  \begin{center}
      \includegraphics[width=3.5cm]{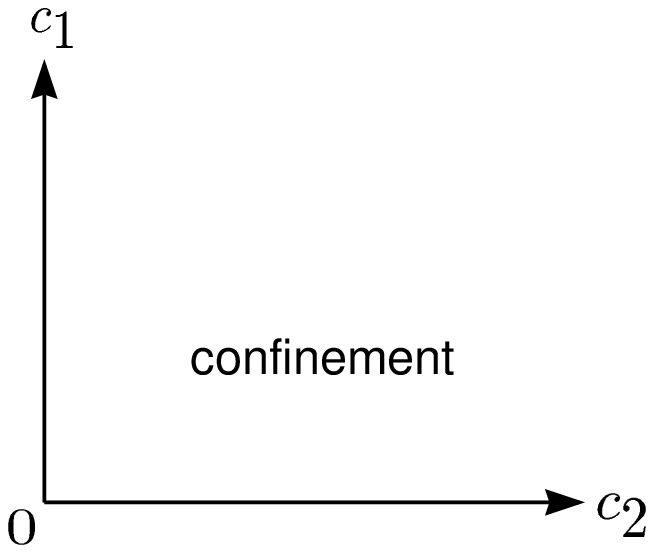}\\
      (c) Compact 3D
  \end{center}
 \end{minipage}
 \begin{minipage}{0.49\hsize}
  \begin{center}
      \includegraphics[width=3.5cm]{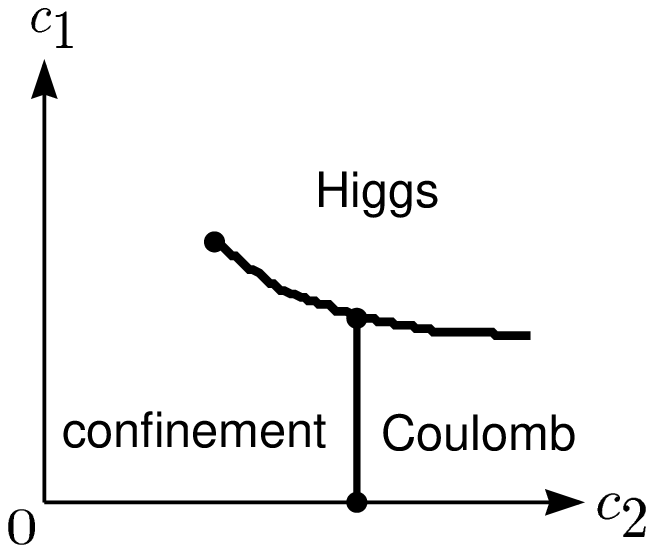}\\
      (d) Compact 4D
  \end{center}
 \end{minipage}
\caption{
Schematic phase diagrams in the $c_2-c_1$plane of the 3D and 4D LGT with compact 
and noncompact 
action. All the Higgs-Coulomb transitions are of second order. 
The Higgs-confinement one in (d) is first order
and the confinemnet-Coulomb one in (d) is weak first order.
The partition function along the line $c_2=0$ is evaluated exactly
by single link integral. It is an analytic function of $c_1$ and so no 
transitions exist along this line.
}
\label{pd}
\end{figure}

In Fig.\ref{pd} we present the schematic phase diagrams in the $c_2-c_1$ plane.
The noncompact 3D and 4D models have the Coulomb phase and the Higgs phase.
The compact 3D model has only the confinement phase.
The compact 4D model has all the three phases.
These points are understood as follows; (i) 
the Higgs phase may exist at large $c_1$ above the Coulomb phase in 3D and 4D,
and above the confinement phase in 4D, (ii) the Coulomb phase may exist
at sufficiently large $c_2$ in 4D case, (iii) the confinement phase exists
in the compact case, but not in the noncompact phase.

We note that, as $c_2$ decreases, all the existing confinement-Higgs 
transition curves terminate at points with $c_2 >0$. This can be understood 
by calculating $Z$ for $c_2=0$ exactly.
Because the integration over $\theta_{x\mu}$ there is factorized as
\be
\hspace{-0.5cm}
Z(c_1,c_2=0) =\prod_{x,\mu}\int d\theta'_{x\mu}\ e^{c_1\cos\theta'_{x\mu}}
=I_0(c_1)^{N_\ell}, \ \theta'_{x\mu}\equiv \varphi_{x+\mu}+\theta_{x\mu}
-\varphi_{x},
\ee
where $I_0(c_1)$ is the modified Bessel function and $N_\ell (\equiv
\sum_{x,\mu}1)$ is the total number of links.
$Z(c_1,c_2=0)$ is then  an analytic function of $c_1$ and no phase transition
can exists along the $c_1$ axis. This fact is known as complementarity
and implies that the confinement phase and the Higgs phase are connected with
each other without crossing a phase boundary. 

Another limiting case $c_2\to \infty$
may imply $\Delta\theta \to 0$. 
Therefore the resulting system becomes
\be
Z(c_1, c_2\to\infty)=\prod_{x,\mu}\int d\varphi'_x 
\exp[c_1\sum_{x,\mu}\cos(\varphi'_{x+\mu}-\varphi'_x)],
\ee
by setting as $\theta_{x\mu}=\nabla_\mu\Lambda_x$ (pure gauge configuration)
with $\varphi'_x \equiv \varphi_x +\Lambda_x$. This system is just the 3D and 4D
XY spin models with the XY spin $(\cos\varphi_x,\sin\varphi_x)$,
which are known to exhibit a second-order transition. This
corresponds to the Higgs-Coulomb transition in Fig.\ref{pd}.

Let us discuss a general correlation function $\la O(\{\theta\}, \{\varphi\})\ra$
(the expectation value of the function $ O(\{\theta\}, \{\varphi\})$)
 of the model (\ref{u1lgt}) defined as
\be
\la O(\{\theta\}, \{\varphi\}) \ra =\frac{1}{Z}
\int[d\varphi][d\theta]O(\{\theta\}, \{\varphi\})\exp(A).
\ee

First, $\la O(\{\theta\}, \{\varphi\}) \ra$ is known to satisfy the Elitzur 
theorem\cite{elitzur}, which says
\be
\la O(\{\theta\}, \{\varphi\}) \ra = 0 \ {\rm if}\
O(\{\theta\}, \{\varphi\})\ {\rm is\ not\ gauge\ invariant.}
\ee
Its proof is rather simple and given by making a change of variables
in a form of gauge transformation and noting the measure 
$[d\theta][d\varphi]$ is invariant under it.
A famous example of gauge-invariant correlation function 
is the Wilson loop defined on a closed loop $C$ along
successive links as
\be
W(C) =\la \prod_{\ell\in C}U_{\ell} \ra.
\ee
It is used by Wilson as a nonlocal ``order parameter" to distinguish a 
confinement phase and a deconfinement phase for pure gauge system
(i.e., involving no dynamical charged particles).
If $W(C)$ satisfies the ``area law" as $C$ becomes large, 
$W(C) \sim \exp(-a S(C))$ where $a > 0$ and $S(C)$ is the minimum area
bound by $C$, the system is in the confinement phase. On the other hand,
if it satisfies the perimeter law, $W(C) \sim \exp(-a' L(C))$
where $L(C)$ is the length of $C$, it is in the deconfinement phase.  
This interpretation comes from the fact that for the rectangular
$C$ of the horizontal size $R$ and the vertical size $T$
[The coordinate $(x_0,x_1)$  of four corners in the 0-1 plane, e.g., are
(0,0), (0,R), (T,0), (T,R)], 
$V(R)$ defined through $W(C)\sim \exp(-V(R)T)$ at sufficiently large $T$ 
expresses the lowest potential
energy stored by a pair of external charges separated by the distance $R$.
The area law implies the confining potential $V(R) =aR$.

Second, one may estimate $Z$ and the correlation functions of 
gauge-invariant objects by fixing the gauge because they are gauge invariant.
Fixing implies to fix one degree of freedom per site $x$ corresponding to
$\Lambda_x$ whatever one wants. For example,
the choice $\varphi_x=0$ is called the unitary gauge, and another choice
$\theta_{x0} =0$ is called the temporal gauge.
In the mean field theory for the compact LGT, one choose gauge variant quantities
$\la \exp(i\theta_{x\mu})\ra$ and $\la \exp(i\varphi_x)\ra$ as 
a set of order parameters and predicts the phase structure involving
the above mentioned three phases.
They should be considered as a result after gauge fixing.
If one wants to make it compatible with the Elitzur theorem, one simply need
to superpose over the gauge copies of these order parameters
as Drouffe suggested\cite{drouffe}. For $\la \exp(i\theta_{x\mu})\ra$, it implies
\be
\prod_x\int_0^{2\pi} \frac{d\Lambda_x}{2\pi} \cdot \la \exp(i\theta'_{x\mu})\ra = 0,\ \theta'_{x\mu}
=\theta_{x\mu} -\nabla_\mu\Lambda_x,
\ee 
without changing the gauge-invariant results such as phase diagram, etc..

\section{Ferromagnetic Superconductivity and its Lattice Gauge Model}
\setcounter{equation}{0}

In this section and Sect.4, we shall consider a theoretical model for the FMSC,
which can be regarded as a kind of LGT.
In some materials such as UGe$_2$, it has been observed that 
SC can coexist with a FM long-range order\cite{fmscexp}.
In UGe$_2$ and URhGe,
a SC appears only within the FM state in the pressure-temperature
($P$-$T$) phase diagram, whereas,
in UCoGe, the SC exists both in the FM and 
paramagnetic states.
Soon after their discovery, phenomenological models of FMSC materials were 
proposed\cite{fmscgl}. In those studies, the FMSC state is  
characterized as a spin-triplet $p$-wave state of electron pairs
as suggested experimentally\cite{pw}. 
In this section, we introduce the GL theory of FMSC materials defined on the
3d spatial lattice (4D Euclidean lattice). As we see, 
the FM order naturally induces nontrivial vector potential (gauge field)
and it competes with another order, the SC order, 
therefore a nonperturbative study is 
required in order to clarify the phase diagram etc.

\subsection{Ginzburg Landau theory of FMSC in the continuum space}

Before considering a lattice model of FMSC itself, 
let us start with the GL theory of FMSC in the 3d continuum space\cite{fmscgl}.
It was introduced phenomenologically  for the case with
strong spin-orbit coupling, although the origin of SC, etc. 
had not been clarified.
Its free energy density $f_{\rm GL}$ contains a pair of basic 3d-vector fields,
$\vec{\psi}(r)$ and $\vec{m}(r)$, which are to be regarded as 
c-numbers appearing in the path-integral expression for the
partition function of the underlying quantum theory as in Eq.(\ref{nrlgt})
[We discuss the partition function in detail later; see Eq.(\ref{Zfmsc})]. 

The three-component complex field $\vec{\psi}=
(\psi_{1},\psi_{2},\psi_{3})^{\rm t}$
is the SC order parameter, 
describing the degrees of freedom of electrons participating in SC.
More explicitly, 
$\vec{\psi}(r)$ describes Cooper-pairs of electrons
with total spin $s=1$ (triplet) and the relative angular momentum
$\ell =1$ ($p$-wave) (Note that only odd $\ell$'s are allowe for $s=1$
by Pauli principle).
The complex Cooper-pair field $\Phi_q(r)$ for a
$s=1$ and $\ell=1$ state generally has $3\times 3=9$ components ($q=1, \cdots, 9$). 
The strong spin-orbit coupling forces the spin $\vec{S}(r)$
and the angular momentum $\vec{L}(r)$ made of $\Phi_q(r)$ parallel each other,
and reduces their degrees of freedom down to 3.
Such a Cooper pair is directly expressed in terms of
the amplitudes defined as
\be
\Delta_{\sigma\sigma'}(r)\equiv 
\int dr_0\Gamma({r_0})\la \hat{C}_\sigma(r+\frac{r_0}{2})\hat{C}_{\sigma'}
(r-\frac{r_0}{2})\ra.
\ee 
$\hat{C}_\sigma(r)$ is the annihilation operator of electron
at $r$ and spin $\sigma = \ua, \da$. $\Gamma(r_0)$ with $r_0$ being the relative 
coordinate of electrons is the weight
reflecting the attractive interaction and the $p$-wave nature.
Because we assign $\vec{\psi}(r)$ as a 3d vector, it is convenient to use the  
so called $\vec{d}$-vector, which transforms also  as a vector in the 3d space;
\be
\vec{\psi}(r)\propto \vec{d}(r)=\left(
\begin{array}{c}
d_x(r)\\
d_y(r)\\
d_z(r)
\end{array}
\right)
\equiv
\left(
\begin{array}{c}
-\frac{1}{2}(\Delta_{\ua\ua}(r)-\Delta_{\da\da}(r))\\
-\frac{i}{2}(\Delta_{\ua\ua}(r)+\Delta_{\da\da}(r))\\
\Delta_{\uparrow\downarrow}(r) (=\Delta_{\downarrow\uparrow}(r))
\end{array}
\right).
\ee 

The real vector field $\vec{m}(r)$ is  the FM order parameter,
describing the degrees of freedom
of electrons participating in the normal state. More explicitly,
it describes  their
magnetization,
\be
\vec{m}(r) =\la \hat{C}'_{\sigma}(r)\Big(-i\delta_{\sigma\sigma'}\vec{r}\times
\vec{\nabla}+\frac{1}{2}\vec{\sigma}_{\sigma\sigma'}\Big)\hat{C}'_{\sigma'}(r)\ra,
\ee
where $\hat{C}'_{\sigma}(r)$ denotes the annihilation operator of electrons
{\it not} participating in the SC state.
Because there holds div $\vec{m}(r) = 0$ as in the usual magnetic field, 
$\vec{m}(r)$ is expressed by using the vector potential $\vec{A}(r)$ 
(gauge field) as
\be
\vec{m}(r)={\rm rot} \vec{A}(r).
\ee
Therefore, the fundamental fields may be $\vec{\psi}(r)$  and $\vec{A}(r)$,

The GL free energy density $f_{\rm GL}$ is then given by\cite{fqhegl}
\begin{eqnarray}
f_{\rm GL}&=&f_{\psi}+f_{m}+f_{\rm Z},\nn
f_{\psi}&=&K\sum_{i}(D_i\vec{\psi})^\ast\cdot (D_i\vec{\psi})+
\alpha(T-T_{\rm{\tiny SC}}^0)\vec{\psi}^\ast\cdot\vec{\psi}
+\lambda(\vec{\psi}^\ast\cdot\vec{\psi})^2,\nn
f_{m}&=&K'\sum_{i}(\partial_i \vec{m})^2+\alpha'(T-T_{\rm{\tiny FM}}^0)\vec{m}^2
+\lambda'(\vec{m}^2)^2,\nn
f_{\rm Z}&=&-J \vec{m}\cdot\vec{S},\
D_i=\partial_i-2ieA_i,\
\vec{S}=-i\vec{\psi}^\ast\times \vec{\psi}.
\label{GL}
\end{eqnarray}
$K^{(')}, \alpha^{(')}, \lambda^{(')}$ are GL parameters;
real positive parameters characterizing each material.
$\vec{S}(r)$  is a real vector field describing the spin of Cooper pairs.
For example, $S_3=\frac{1}{2}(|\Delta_{\ua\ua}|^2-|\Delta_{\da\da}|^2)$. 
$f_{\rm Z}$ with $J > 0$ is nothing but the Zeeman coupling of 
$\vec{S}$ and $\vec{m}$, which enhances
the coexistence of the FM and SC orders as one easily expects.
Although the existence of $f_{\rm Z}$ term is supported phenomenologically, 
its microscopic origin 
should be clarified by detailed study of the interactions
of electrons (and nuclei) in each material.

$f_{\rm GL}$ is gauge invariant under the following U(1) gauge transformation,
\be
\vec{\psi}(r)&\to&\vec{\psi}'(r)=e^{i\lambda(r)}\vec{\psi}(r),\
\vec{A}(r)\to\vec{A}'(r)=\vec{A}(r)+\frac{1}{2e}\vec{\nabla}\lambda(r).
\ee
So $\vec{m}(r)$ and $\vec{S}(r)$ are gauge invariant.
This gauge invariance reflects the original U(1) gauge invariance of
the EM interaction of electrons.
For example, the $K$ term with the covariant derivative $D_i$
describes the interaction of SC Cooper pairs of 
charge $-2e$ with the vector potential $\vec{A}$ made of the
normal electrons. Its microscopic origin  should be traced back to the repulsive 
Coulombic interactions between electrons.

Before introducing the lattice GL theory, let us list up some main 
characteristics suggested by the continuum GL theory (\ref{GL}).
\begin{itemize}
\item
In the mean-field approximation with ignoring the Zeeman coupling
$f_Z$, $T^0_{\rm FM}$ and $T^0_{\rm SC}$ are critical temperatures
of the FM and SC phase transitions, respectively.
\item
Existence of a finite magnetization, $\la \vec{m} \ra \neq 0$,
 means a nontrivial configuration of $A_i(r)$. This 
 induces nontrivial spatial dependence  of 
the Cooper pair $\vec{\psi}$ through
the kinetic term $K(D_i\vec{\psi})^\ast\cdot (D_i\vec{\psi})$ in $f_{\rm GL}$,
because this term   favors $D_i\vec{\psi}\simeq 0$.
A typical example of such configurations is a vortex configuration
characterized by a nonvanishing vorticity $\vec{v}(r)$ of $\vec{\psi}$.
The 3d vector $\vec{v}(r)$ of a complex field $\phi(r)=|\phi(r)|\exp(i\varphi(r))$
is defined generally as the winding number of its phase $\varphi(r)$. 
Explicitly, the component of $\vec{v}(r)$ along the direction of a normal vector 
$\vec{n}$ is defined by the circle integral over $\varphi(r)$, 
\be
\vec{n}\cdot\vec{v}(r') = \frac{1}{2\pi}\oint_{C_{\vec{n}}(r')} 
d\varphi(r)\ (\in \bf{Z}),
\ee
where $C_{\vec{n}}(r')$ is a circle  lying in the plane 
parpendicular  to $\vec{n}$ with its center at $r'$.
$\vec{n}\cdot\vec{v}(r)$ takes integers due to the single-valuedness of $\phi(r)$.
In the later sections, we shall consider two vorticities $\vec{v}^{\pm}(r)$ 
corresponding to $\psi_1(r)\pm i\psi_2(r)$ respectively.
Then the simple mean-field like estimation of the SC critical temperature
has to be reexamined by more reliable analysis.
\item The Zeeman coupling $f_{\rm Z}$ obviously enhances the coexisting
phase of the FM and SC orders, since  it favors a set of antiparallel 
$\vec{S}$ and $\vec{m}$ with large $|\vec{\psi}|$ and $|\vec{m}|$.
\item
In the FM phase, $\vec{A}$ produces a nonvanishing magnetic field
$\vec{m}(r)$ inside the materials.
Because the $K$ term in $f_{\psi}$ of Eq.(\ref{GL})
disfavors the coherent (spatially uniform) condensation 
of $\vec{\psi}$ ($\partial_i\vec{\psi}=0$),
 due to rot ${A} \neq 0$
(favoring nontrivial ones such as vortices as explained above),
the SC in a magnetic field is disfavored.
In this sense, the system (\ref{GL}) is a kind of a frustrated system
(the $K$ term disfavors coexistence although the Zeeman term does).
\end{itemize}

Then a detailed study by using numerical methods is needed to obtain the 
correct phase diagram.
To this end, we shall introduce a lattice version of the GL theory 
(\ref{GL})
that is suitable for the investigation by using MC simulations. 

\subsection{GL theory on the lattice}

In this subsection we explain the lattice GL theory introduced
in Ref.\cite{shimizu}.
In constructing the theory, 
we start with the following two simplifications: 
\begin{enumerate}
\item We consider the ``London" limit of the SC, i.e., we assume 
$\vec{\psi}^\ast(r)\cdot\vec{\psi}(r)={\rm const.}$ ignoring the fluctuations
of the radial degrees of freedom $|\vec{\psi}(r)|$. 
This assumption is legitimate as the phase degrees of freedom $\vec{\psi}(r)$
themselves play an essentially important role in the SC transition
(Recall the account of phase coherence for a Bose-Einstein condensate)
\cite{radialcomp}. 
\item 
We assume that the third component 
$\psi_3(r)\propto \Delta_{\uparrow\downarrow}(r)$ is negligibly small 
compared to the remaining ones $\psi_{1,2}(r)$,
\be
\hspace{-1cm}
\psi_\pm\equiv
\displaystyle{\frac{1}{\sqrt{\mathstrut 2}}} 
(\psi_1\pm i\psi_2),\
\psi_{\ua\ua}\equiv \psi_- \propto \Delta_{\ua\ua},\
\psi_{\da\da}\equiv -\psi_+ \propto \Delta_{\da\da}.
\ee
This simplification is consistent with the fact that 
the real materials exhibit FM orders of Ising-type
with the $i=3$-direction as the easy axis.
In fact, $\vec{S}=-i\vec{\psi}^\ast\times\vec{\psi}$ 
is calculated with $\psi_{\ua\da}\equiv \psi_3$ as
\be
\hspace{-1cm}
\vec{S}=(-\sqrt{\mathstrut 2}\ {\rm Re}(\psi_{\uparrow\uparrow}+
\psi_{\downarrow\downarrow})\psi_{\uparrow\downarrow}^*,
\sqrt{\mathstrut 2}\ {\rm Im}(\psi_{\uparrow\uparrow}-
\psi_{\downarrow\downarrow})\psi_{\uparrow\downarrow}^*,
|\psi_{\uparrow\uparrow}|^2-|\psi_{\downarrow\downarrow}|^2)^{\rm t}.
\ee
So the Zeeman coupling $f_{\rm Z}$ requires large $\psi_{1,2}$  
compared to $\psi_3$ to favor $\vec{m} \propto (0,0,m)$. 
\end{enumerate}
In summary, we parametrize the Cooper-pair field in the London limit, $|\psi(r)|^2
=[\alpha/(2\lambda)](T^0_{\rm {\tiny SC}}-T)$, at low $T (< T^0_{\rm {\tiny SC}})$ 
with the third easy axis as
\be
\left(
\begin{array}{c}
\psi_1(r)\\
\psi_2(r)\\
\psi_3(r)
\end{array}
\right)
=
\sqrt{{\alpha \over 2\lambda}(T^0_{\rm {\tiny SC}}-T)}\times
\left(
\begin{array}{c}
z_1(r)\\
z_2(r)\\
0
\end{array}
\right),\ |z_1(r)|^2+|z_2(r)|^2=1.
\label{CP1}
\ee
The ``normalized" two-component complex field $z_a(r)\ (a=1,2)$
satisfying the constraint
(\ref{CP1}) is called CP$^1$ variable (CP stands for complex projective group).
In terms of the SC order-parameter field $z_a(r)$, 
the first $K$ term of (\ref{GL}) is rewritten as
$K\alpha(2\lambda)^{-1}(T^0_{\rm {\tiny SC}}-T)
\sum_i\sum_a\overline{D_i z_a}\cdot D_i z_a$.

Now let us introduce the lattice GL theory of FMSC defined on the 3d 
cubic lattice. Its free energy density per site $f_r$ is given by\cite{shimizu} 
\begin{eqnarray}
f_r &=& -\frac{c_1}{2}\sum_{i=1}^3\sum_{a=1}^2\left( 
\bar{z}_{r+i,a}\bar{U}_{ri}z_{ra}+ {\rm c.c.}\right)-
c_2 \vec{m}_r^2-c_3 \vec{m}_r\cdot\!\vec{S}_r +c_4(\vec{m}_r^2)^2\nn 
&&-c_5  \sum_i\vec{m}_{r+i}\cdot\vec{m}_r, \
\sum_{a=1}^2 \bar{z}_{ra} z_{ra}=1,\ U_{ri} \equiv \exp(i\theta_{ri}).
\label{FLGL}
\end{eqnarray}
The five coefficients $c_i$ $(i=1\sim 5)$ in (\ref{FLGL})
 are real nonnegative parameters 
that are to distinguish various materials in various environments.
$z_{ra}$ is the CP$^1$  variable put on the site $r$ and 
plays the role of SC order-parameter field. 
$U_{ri}$ is the exponentiated vector potential, $\theta_{ri}$, 
put on the link ($r,r+i$). 
$\vec{m}_r=(m_{r1},m_{r2},m_{r3})^{\rm t}$ is 
the magnetic field made out of $\theta_{ri}$ as
\begin{eqnarray}
\hspace{-0.8cm}
m_{ri} \equiv\! \sum_{j,k=1}^3
 \epsilon_{ijk}\nabla_j \theta_{rk}, \
 \nabla_j \theta_{rk} \equiv \theta_{r+j,k}\!-\!\theta_{rk}.
\label{rota}
\end{eqnarray}
$\vec{m}_r$ serves as the FM order-parameter field. 
Following Eq.(\ref{GL}), the spin $\vec{S}_r$ of the Cooper pair is defined as 
\be
\vec{S}_r=(0,0,S_{r3}), \;\;
S_{r3} \equiv -i (\bar{z}_{r1} z_{r2}-\bar{z}_{r2} z_{r1})\
(\propto |\psi_{\uparrow\uparrow}|^2-|
\psi_{\downarrow\downarrow}|^2),
\ee
where we absorbed the normalization of $\vec{S}(r)$ into the coefficient
$c_3$.
$\vec{m}_r$, $\vec{S}_r$ and
$f_r$ are invariant under the following gauge transformation,
\be
z_{ra} &\rightarrow& z_{ra}'=e^{i\lambda_r} z_{ra},\
U_{ri}\rightarrow U_{ri}'=
e^{-i\lambda_{r+i}}\ U_{ri}\ e^{i\lambda_r}.
\label{gaugesym}
\ee

Here we comment on the way of putting a 3d continuum vector field 
on the 3d lattice.
It sounds natural that a vector field $\vec{B}(r)$ 
should be put on the link $(r,r+i)$ as $B_{ri}$.  
However, it is too naive. In fact, the gauge field $\vec{A}(r)$ has 
the nature of connection as explained and the connection between
two points separated by finite distance such as nearest-neighbor 
pair of sites cannot be implemented by $\theta_{ri}$ itself
but by its exponentiated form such as $U_{ri}$.
This is the reason why one has $\bar{z}_{r+i,a}\bar{U}_{ri}z_{ra}$ term 
in Eq.(\ref{FLGL}). 
On the other hands, concerning to the SC order field 
$\vec{\psi}(r)$,
its suitable lattice version depends on the coherence length $\xi$.
If $\xi$ is of the same order of lattice spacing $a$, its vector nature
should be respected and the link field $\psi_{ri}$ is adequate.
If $\xi \gg a$, the detailed lattice structure is irrelevant and
a simpler version $\psi_{ra}\ (a=1,2,3)$ on the site 
(or its Ising counterpart $z_{ra}\ (a=1,2)$) will do.  Eq.(\ref{FLGL}) 
corresponds to the latter case $\xi \gg a$ as some materials show. 

Let us see the meaning of each term in $f_r$.
The $c_1$-term describes a hopping of Cooper pairs from site 
$r$ to $r+i$ (and from $r+i$ to $r$).
As the Cooper pair has the electric charge $-2e$, it minimally
couples with the vector potential $\theta_{ri}$ via $U_{ri}$
as explained above.
It is important to observe the relation between
the hopping parameter $c_1$ and $T$.  From the expression
given in the paragraph below Eq. (\ref{GL}), we have
\be
c_1 \sim K\alpha \lambda^{-1}(T^0_{\rm {\tiny SC}}-T)a.
\label{c1}
\ee
At sufficiently large $\beta c_1=c_1/T$ (we set $k_{\rm B}=1$) 
that corresponds to low $T$'s, 
the $c_1$-term stabilizes the combination of phases of $\bar{z}_{r+i,a}
\bar{U}_{ri}z_{ra}$, and then, if $U_{ri}$ is stabilized already by other 
mechanism, a {\em coherent condensation of the phase degrees of freedom of}  
$z_r$ {\it is realized inducing the superconductivity}.
The $c_2$ and $c_4$-terms are the quartic GL potential of 
$\vec{m}_r$, favoring a finite amount of {\it local} magnetization
$\la \vec{m}_r \ra \neq 0$ (note that we take $c_2 > 0$). 
We should remark that these terms controls intrinsic magnetization,
which  is different from the fluctuating but external magnetic field.
The $c_5$-term represents the NN coupling of the magnetization
and corresponds to the $K'$ term of Eq.(\ref{GL}). It
enhances uniform configurations of 
$\vec{m}_r$, i.e.,  a FM long-range order signaled by
a finite magnetization $m$,
$m^2\equiv \lim_{|r-r'|\to \infty}\la\vec{m}_r\cdot\vec{m}_{r'}\ra \neq 0$.
The $c_3$-term is the Zeeman coupling, which favors 
collinear configurations of $\vec{m}_r$ and $\vec{S}_r$, 
i.e., enhances the coexistence of FM and SC orders.

The partition function $Z$ at $T$ 
for the energy density $f_r$ of Eq.(\ref{FLGL}) is given by
the integral over a set of two fundamental fields $z_{ra}$ and  $\theta_{ri}$ as
\be
Z &=& \int[dz][d\theta]\exp(-\beta F),\ \beta=T^{-1},\ F=\sum_r f_r,\nn 
\left[dz\right]&=&\prod_r d^2z_{r1}d^2z_{r2}\ \delta(|z_{r1}|^2+|z_{r2}|^2-1),
\nn
\left[d\theta\right]&=&G(\{\theta\})\prod_{r,i} d\theta_{ri},\ \theta_{ri} 
\in (-\infty,\infty). 
\label{Zfmsc}
\ee
where $G(\{\theta\})$ is a gauge fixing function[see the paragraph containing
Eq.(\ref{gv})].

We stress here that the free energy $F$ and the integration variables
in Eq.(\ref{Zfmsc}) have no dependence on the imaginary time $x_0$ in contrast with 
the expression (\ref{u1lgt}). 
The ordinary GL theory explained in the literature also shares this properties.
This implies
that we ignore the $x_0$-dependent modes of would-be $\varphi_x$ and 
$\theta_{x\mu}$ as $A(\{\varphi_x\},\{\theta_{x\mu}\})
\to \beta F(\{\varphi_r\},\{\theta_{ri}\})$.
This is an approximation applicable for small $\beta$'s,
i.e., for high $T$'s. This procedure allows us to make use of 
MC method that is based on the probabilistic
process. In fact, the form (\ref{Zfmsc}) has a real function $F$ and  the 
Boltzmann factor $\exp(-\beta F)/Z$ is able to be interpreted
as a probability. This is in contrast with the
original action $A$ corresponding to $F$; $A$ is certainly 
a complex function as Eq.(\ref{u1lgt})
giving rise to a complex probability.  
We come back to this point later.

The coefficients $c_i \ (i=1,\cdots, 5)$ in $f_r$ may have nontrivial
$T$-dependence as Eqs.(\ref{GL}) and (\ref{c1}) suggest. However,
in the present study we consider the response of the system 
by varying the ``temperature" $T \equiv 1/\beta$
defined by $\beta$,  
an overall prefactor in Eq.(\ref{Zfmsc}), while keeping $c_i$ fixed.
This method corresponds to well-known studies such as the FM transition
by means of the $O(3)$ nonlinear-$\sigma$ model\cite{sigma} and 
the lattice gauge-Higgs models  discussed in Sec.2.1, and is sufficient to
determine the critical temperature [See, e.g., Eq.(\ref{TTSC2})].

In the following section, we shall show the results obtained by MC numerical
evaluation of Eq.(\ref{Zfmsc}).
Physical quantities ${\cal O}(z,\theta)$ like the internal energy, specific heat, 
correlation functions,
etc. are also calculated by the MC simulations as
\begin{equation}
\langle {\cal O}(z,\theta) \rangle =\frac{1}{Z}\int[dz][d\theta]
{\cal O}(z,\theta) 
\exp(-\beta F).
\label{O}
\end{equation}
From the obtained results, we shall clarify the phase diagram of the system
and physical properties of each phase.

\section{Results of Monte Carlo simulation}
\setcounter{equation}{0}

In the present section, we introduce and discuss the results of numerical 
calculations\cite{shimizu,noguchi} 
for the lattice GL model (\ref{FLGL}), which include the phase diagram
and Meissner effect in the SC phase.
Figs.\ref{C:mag}$\sim$\ref{snapvor1} presented 
below are from Ref.\cite{shimizu}
and  Fig.\ref{figb20}$\sim$\ref{figb52} are from Ref.\cite{noguchi}.

\subsection{Phase diagram and Meissner effect}

We first seek for a suitable boundary condition (BC) of $\theta_{ri}$
as the magnetization 
is expressed by $\theta_{ri}$ as Eq.(\ref{rota}).
A familiar condition is the periodic boundary condition such as $\theta_{r+Li,j}
=\theta_{rj}$ 
for the linear system size $L$. However, this condition 
necessarily gives rise to a vanishing net mean magnetization due to the
lattice Storkes theorem.
So we consider the 3d lattice of the size $(2+L+2)^2\times L$ and
take a ``free BC" on $z_r$ in the $i=1,2$ directions defined by
\be
\hspace{-0.3cm}
z_{r+i,a} - \bar{U}_{ri}z_{ra} = 0\
 {\rm for}\ r=\left\{ 
\begin{array}{ll}
(0,r_2,r_3),& i=1,\\
(L+2,r_2,r_3),& i=1, \\
(r_1,0,r_3),& i=2, \\
(r_1,L+2,r_3),& i=2,
\end{array}\right.
\label{bc1}
\ee
whereas we impose the free boundary condition on $\theta_{ri}$.
It is easily shown that the BC (\ref{bc1}) implies that the suppercurrent
$j^{\rm SC}_{ri}$,
\be
j^{\rm SC}_{ri} \propto {\rm Im}\ (\sum_a \bar{z}_{r+i,a}\bar{U}_{ri} z_{ra}),
\ee
satisfies $j^{\rm SC}_{r1(2)} = 0$ on the boundary surfaces in the $2(1)-3$
planes.
That is, the supercurrent does not leak out of the SC material.
Another possible BC is the one that imposes vanishing 
Cooper-pair amplitude $z_{ra}$, but we expect, as usual, 
that the qualitative bulk properties
of the results are not affected by the BC seriously.
[The above BC (\ref{bc1}) means that the magnetization is just	
like the Ising type and $\langle\vec{m}_x\rangle=(0,0,m_3)$, which
describes the real materials properly.]

In the simulation made in Refs.\cite{shimizu,noguchi} 
we use the standard Metropolis algorithm\cite{metropolis} 
for the lattice size up to $L=30$. The typical number of sweeps 
for a measurement is $(30000\sim 50000)\times 10$ and the acceptance 
ratio is $40\%\sim50\%$. The error is calculated as the standard deviation
of 10 samples obtained by dividing one measurement run into 10 pieces.
Concerning to the gauge-fixing term of Eq.(\ref{gv}),
even for a noncompact gauge theory such as the present one, 
gauge fixing is not necessary in practical MC simulations. 
This is because, in addition to the reasons given
at the end of paragraph of Eq.(\ref{gv}),
the MC updates using random numbers
almost never generate exactly gauge-equivalent configurations.

We calculate the internal energy $U$,
the specific heat $C$ of the central region $R$ of the size $L^3$, 
the magnetization  $m_i$, which are defined as follows,
\be
\hspace{-0.9cm}
U&=&\frac{1}{L^3}\langle F_{L} \rangle,\
C=\frac{1}{L^3}\langle (F_{L}-\langle F_{L} \rangle)^2\rangle,\
F_{L}\equiv \sum_{r \in R} f_r,\ 
m_i\equiv\frac{1}{L^3}\langle |\sum_{r \in R} m_{ri}| \rangle,
\ee
and the normalized correlation functions, 
\be
G_{m}(r-r_0)&=&
\frac{\la\vec{m}_{r}\cdot\vec{m}_{r_0}\ra}{\la\vec{m}_{r_0}\cdot\vec{m}_{r_0}\ra},\
G_{S}(r-r_0)=\frac{\la S_{r3}S_{r_0,3}\ra}{\la S_{r_0,3}S_{r_0,3}\ra},
\label{GmGs}
\ee
where $r_0$ is chosen on the boundary of $R$ such as $(3, 2+L/2, z)$.
Singular behaviors of $U$ and/or $C$ indicate the existence of phase
transitions, and the magnetization $m_i$ and the correlation functions
in Eqs.(\ref{GmGs}) clarify the physical meaning of the observed phase
transitions.
As the present system is a frustrated system as we explained above,
$G_m$ and $G_S$ may exhibit some peculiar behavior in the FMSC state
(the state in which FM and SC orders coexist).

\begin{figure}[b]
 \begin{minipage}{0.45\hsize}
  \begin{center}
  \hspace{-0.6cm}
   \includegraphics[width=4.3cm]{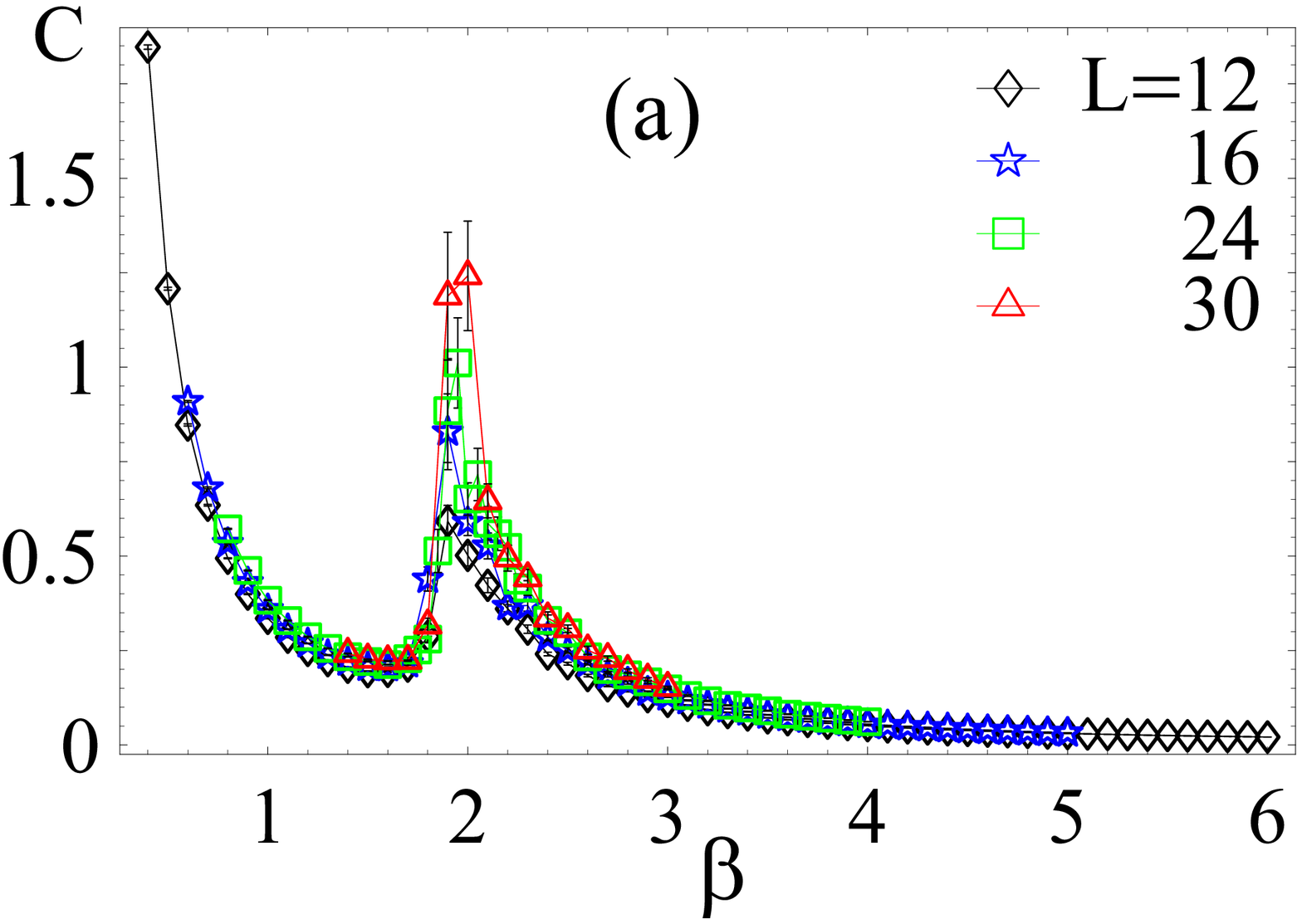}
  \end{center}
 \end{minipage}
     \hspace{-0.6cm}
 \begin{minipage}{0.45\hsize}
  \begin{center}
   \vspace{-0.6cm}
      \includegraphics[width=4.6cm]{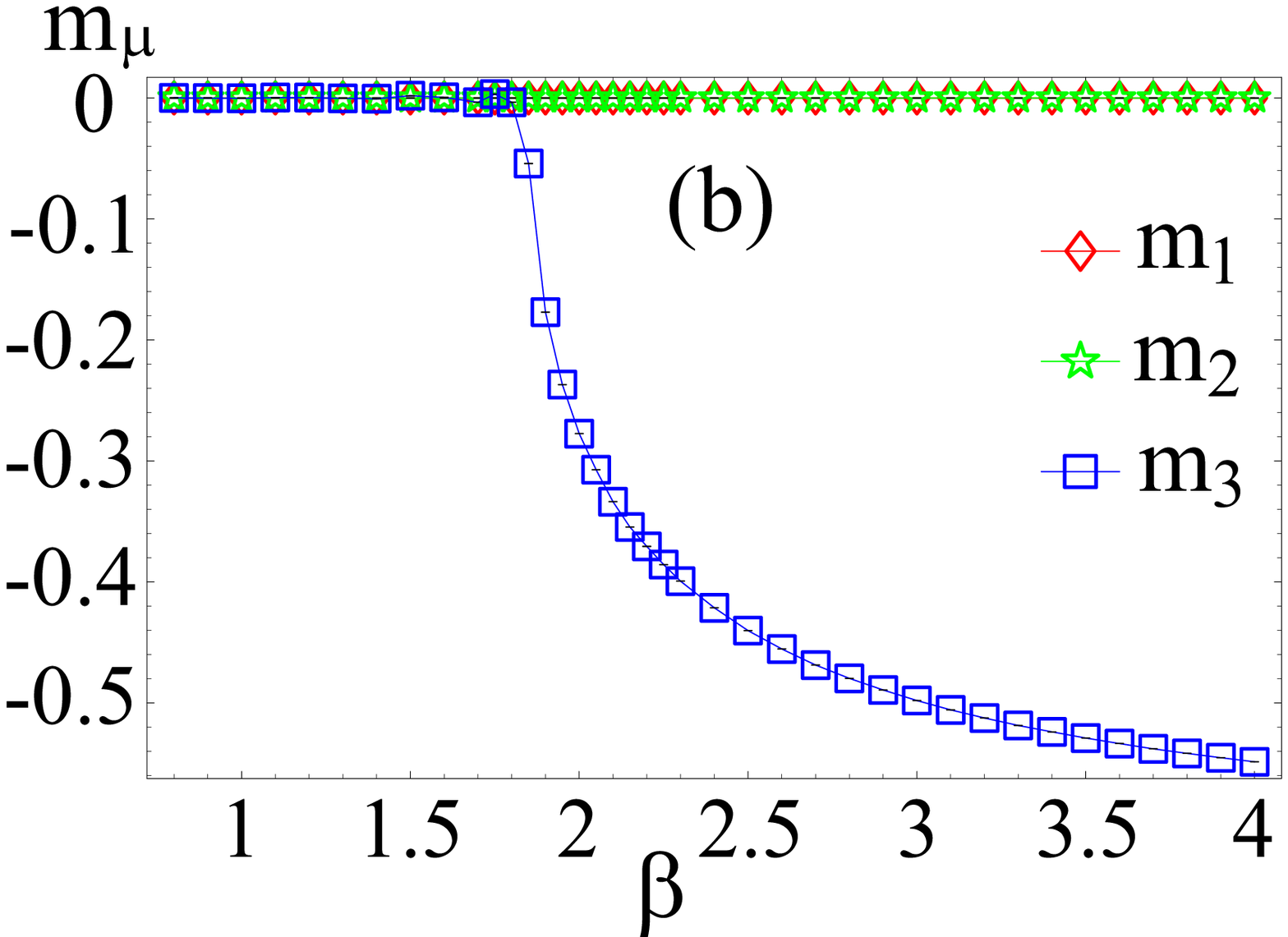}
  \end{center}
 \end{minipage}
\vspace{-0.2cm}
\caption{
(a) Specific heat for
$(c_2,c_4,c_5)=(0.5,4.0,1.0)$ and $c_1=c_3=0$.
At $\beta\simeq 2.0$, $C$ exhibits a sharp peak indicating
a second-order FM phase transition.
(b) Each component of magnetization $m_\mu$
vs $\beta$.
For $T<T_{\rm FM}$, $m_3$ develops, whereas 
$m_1$ and $m_2$ are zero within the errors as expected.
}
\label{C:mag}
\end{figure}

To verify that the model (\ref{Zfmsc}) actually exhibits a FM phase transition
as $T$ is lowered,
we put $c_1=c_3=0$ and $(c_2,c_4,c_5)=(0.5,4.0,1.0)$,
and measured $C$  
increasing $\beta$ in the Boltzmann factor of Eq.(\ref{Zfmsc}). 
In Fig.\ref{C:mag}
we show $C$ and $m_i$, which obviously indicate that a 
second-order phase transition to the FM state takes place
at $\beta_{\rm FM}=1/T_{\rm FM}\simeq 2.0$. 
We observed that other cases with various values of $c_{2,4,5}$  
exhibit similar FM phase transitions.

\begin{figure}[t]
\vspace{-1cm}
\begin{center}
\hspace{-3cm}
\includegraphics[width=9cm]{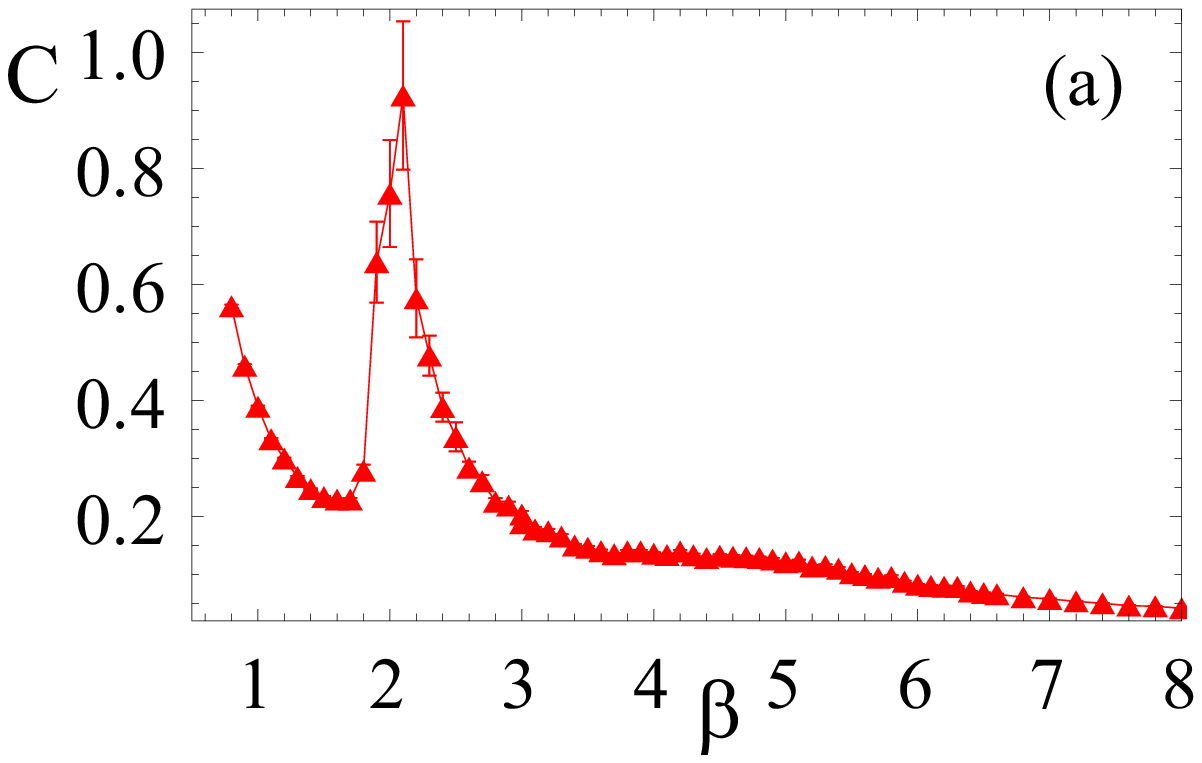}
\end{center}
\vspace{-0.5cm}
 \begin{minipage}{0.45\hsize}
  \begin{center}
  \hspace{-0.6cm}
\includegraphics[width=4.0cm]{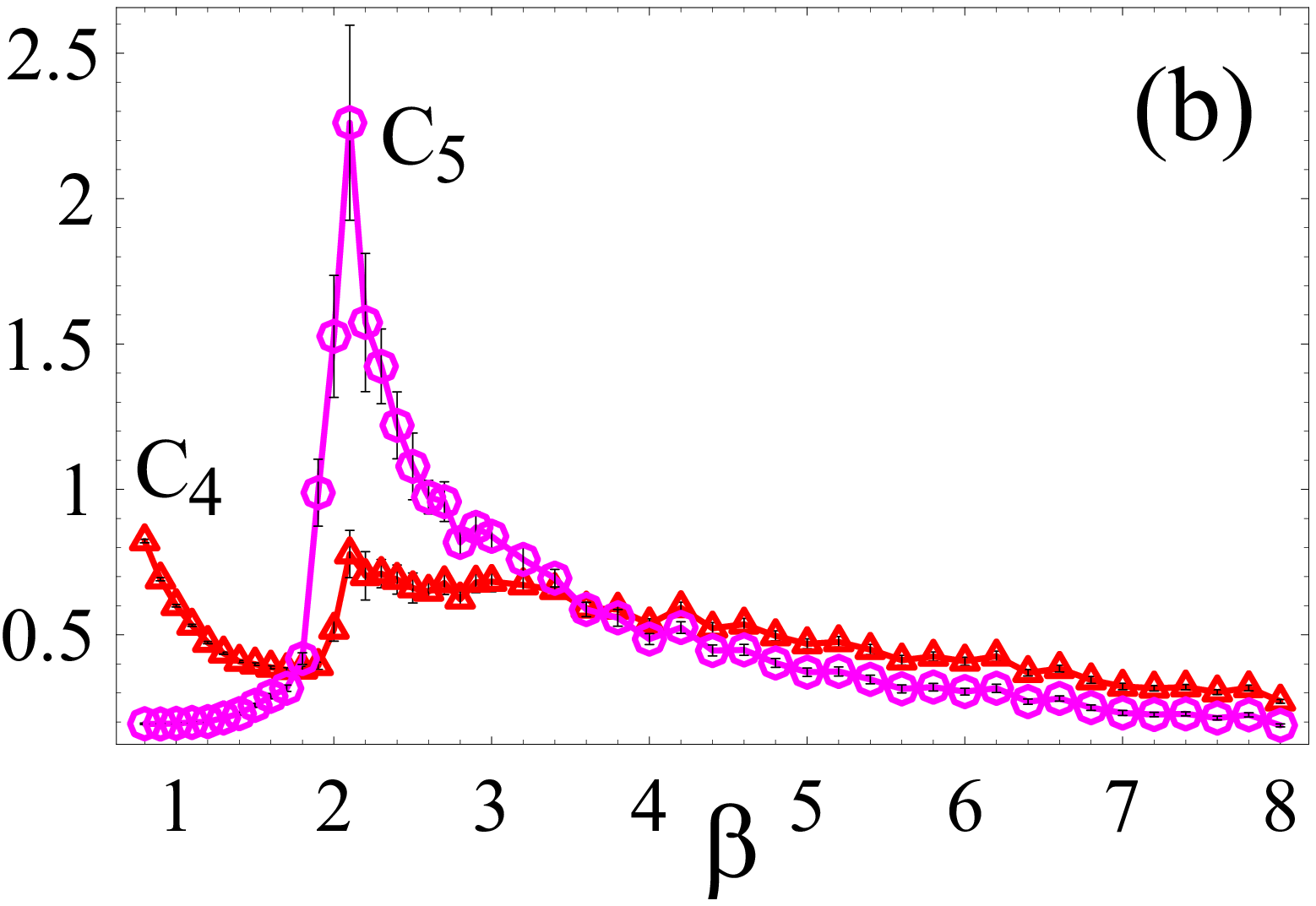}
  \end{center}
 \end{minipage}
     \hspace{-0.2cm}
 \begin{minipage}{0.45\hsize}
  \begin{center}
   \vspace{-0.9cm}
\includegraphics[width=4.2cm]{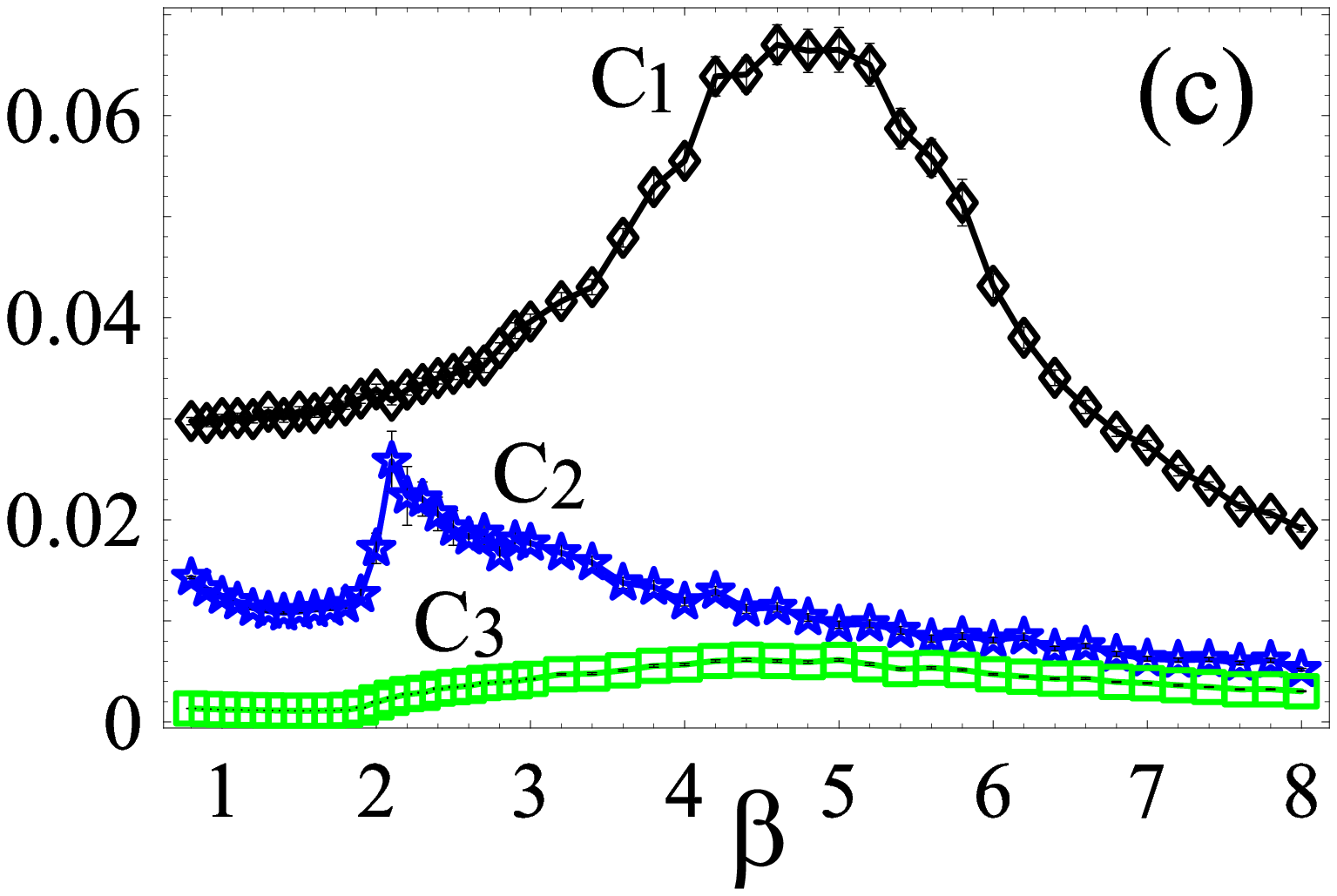}
  \end{center}
 \end{minipage}
 \vspace{-1.0cm}
\caption{
(a) Specific heat vs $\beta$ for $(c_1,c_3)=(0.2,0.2)$ and
$(c_2,c_4,c_5)=(0.5,4.0,1.0)$ ($L=20$). 
There are a large peak at $\beta \simeq 2.1$ and a small one at
$\beta \simeq 4.5$.
(b,c) Partial specific heat $C_i$ of Eq.(\ref{psh}) vs $\beta$.
The small and broad peak at $\beta \simeq 4.5$ in $C$ is related to
fluctuations of $c_1$-term. 
}
\label{C:magsc}
\end{figure}

Let us see the SC phase transition and how it coexists with
the FM.
We recall that the case of all $c_i=0$ except for $c_1$
was studied in the previous paper Ref.\cite{takashima}.
There it was found that the phase transition from the confinement phase 
to the Higgs phase takes place at $c_1 \simeq 2.85$. 
This result suggests that the SC state exists at sufficiently large $c_1$
also in the present system with $c_{i(\neq 1)}\neq 0$.

To study this possibility, we use
$(c_2,c_4,c_5)=(0.5,4.0,1.0)$ as in the FM sector above and
$(c_1,c_3)=(0.2,0.2)$.
In Fig.\ref{C:magsc}a, we show $C$ vs $\beta$, which exhibits
a large and sharp peak at $\beta \simeq 2.1$ and a small and 
broad one at $\beta \simeq 4.5$.
To understand the physical meaning of the second broad peak,
it is useful to measure ``partial specific heat" $C_i$ for 
each term $F_i$ in the
free energy (\ref{FLGL}) defined by 
\be C_i=
\frac{1}{L^3}\la(F_{i}-\la F_{i}\ra)^2\ra, \ F_i=\sum_{r \in R} f_{ir},
\label{psh}
\ee
where $f_{ir}$ is the $c_i$-term in $f_r$ of Eq.(\ref{FLGL}).

Figs.\ref{C:magsc}b,c show that the partial specific heat 
$C_{2,4,5}$ have a sharp peak at $\beta\simeq 2.1$. Thus the
peak of total $C$ there should indicate the FM phase transition.
The analysis of Meissner effect (see the discussion below using Fig.\ref{MG})
supports this point.
On the other hand, $C_1$
of the $c_1$-term has a relatively large and broad peak at
$\beta \simeq 4.5$. $C_3$ also shows a broader peak there.
Then we judge that the SC phase transition takes place
at $\beta_{\rm SC}\simeq 4.5$.
To support these conclusions, we show 
$G_{m}(r)$ and $G_{S}(r)$ in Fig.\ref{correlation1}.
At $\beta=2.5$, $G_{m}(r)$ exhibits a 
finite amount of the FM order, whereas $G_{S}(r)$
decreases very rapidly to vanish.
This means that, as $T$ is decreased, 
the FM transition takes place first and then
the SC transition does.
Therefore, for $\beta\geq\beta_{\rm SC}\simeq 4.5$, the FM and SC 
orders coexist.
In this way, the partial specific heat may be used to judge the nature
of each transition found by the peak of full specific heat $C$
(We note $C \neq \sum_i C_i$ in general due to interference). 

\begin{figure}[b]
\begin{center}
\hspace{-0.3cm}
\includegraphics[width=4.3cm]{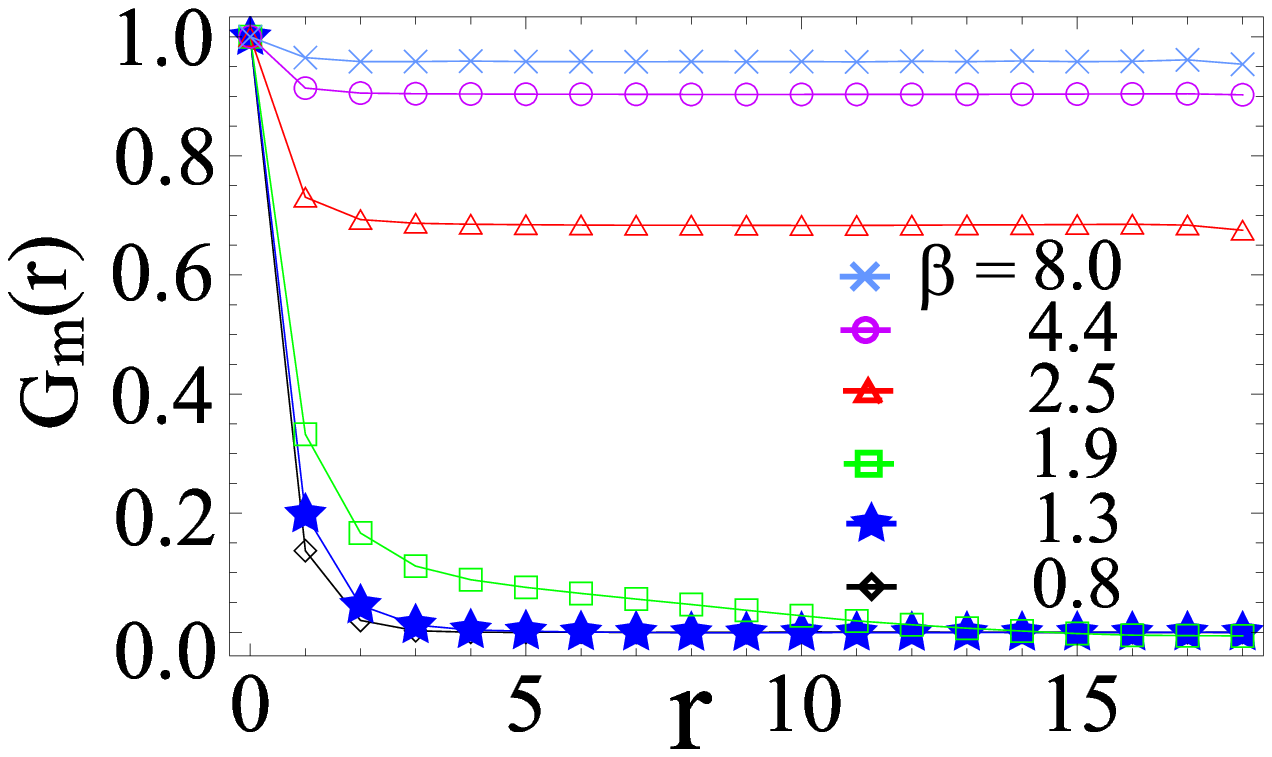}
\hspace{-0.0cm}
\includegraphics[width=4.3cm]{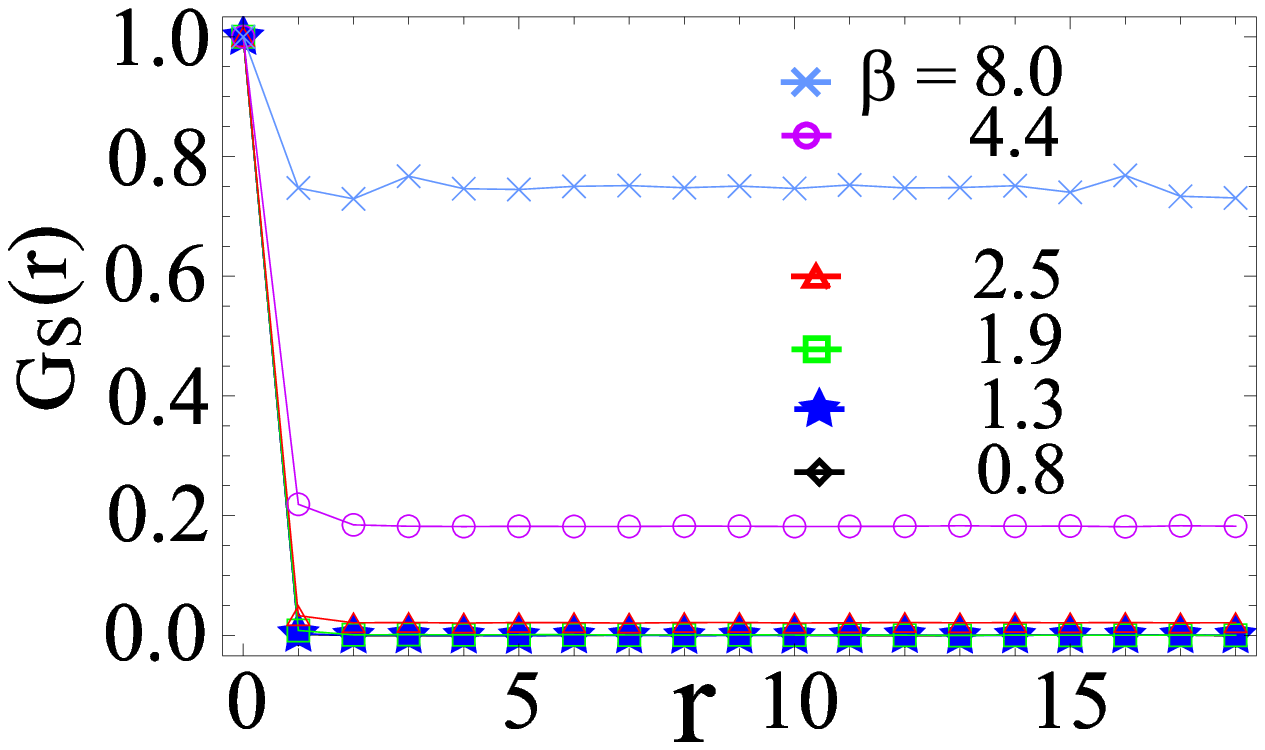}
\end{center}
\vspace{-0.3cm}
\caption{
Correlation functions $G_{m}(r)$ and $G_{S}(r)$ at various $T$'s
for $L=20$.
$c_i$'s are the same as in Fig.\ref{C:magsc}. 
}
\label{correlation1}
\end{figure}

As we explained in the previous section, 
the bare transition temperature $T^0_{\rm SC}$ in Eq.(\ref{GL}) and the genuine 
transition temperature $T_{\rm SC}$ are different.
Then it is interesting to clarify the relation between them.
From Eq.(\ref{Zfmsc}), any physical quantity is a function of $\beta c_i$.
In the numerical simulations, we fix the values of $c_i$ and vary $\beta$
as explained.
Then the result $\beta_{\rm SC} \simeq 4.5$ means
\be
\beta c_1|_{T=T_{\rm SC}}=4.5\times 0.2.
\label{TTSC}
\ee
By using Eq.(\ref{c1}), this gives the following relation;
\be
&&{1 \over T_{\rm SC}}{K\alpha (T^0_{\rm SC}-T_{\rm SC}) a\over \lambda}
=0.90\ \to
T_{\rm SC}=\Big(1+{0.90\ \lambda \over K\alpha\ a}\Big)^{-1}T^0_{\rm SC}.
\label{TTSC2}
\ee
Eq.(\ref{TTSC2}) shows that the transition temperature
is lowered by the fluctuations of the phase degrees of freedom of
Cooper pairs.
We expect that a relevant contribution to lowering the SC transition 
temperature comes from vortices that are generated spontaneously 
in the FMSC as we shall show in Sec.4.2.

After having confirmed that the genuine critical temperature can be
calculated by the critical value of $\beta$ with fixed $c_i$, 
we use the word temperature in the rest of the paper just as the one defined 
by $T\equiv 1/\beta$ while $c_i$ are $T$-independent parameters.

\begin{figure}[t]
\begin{center}
\includegraphics[width=5cm]{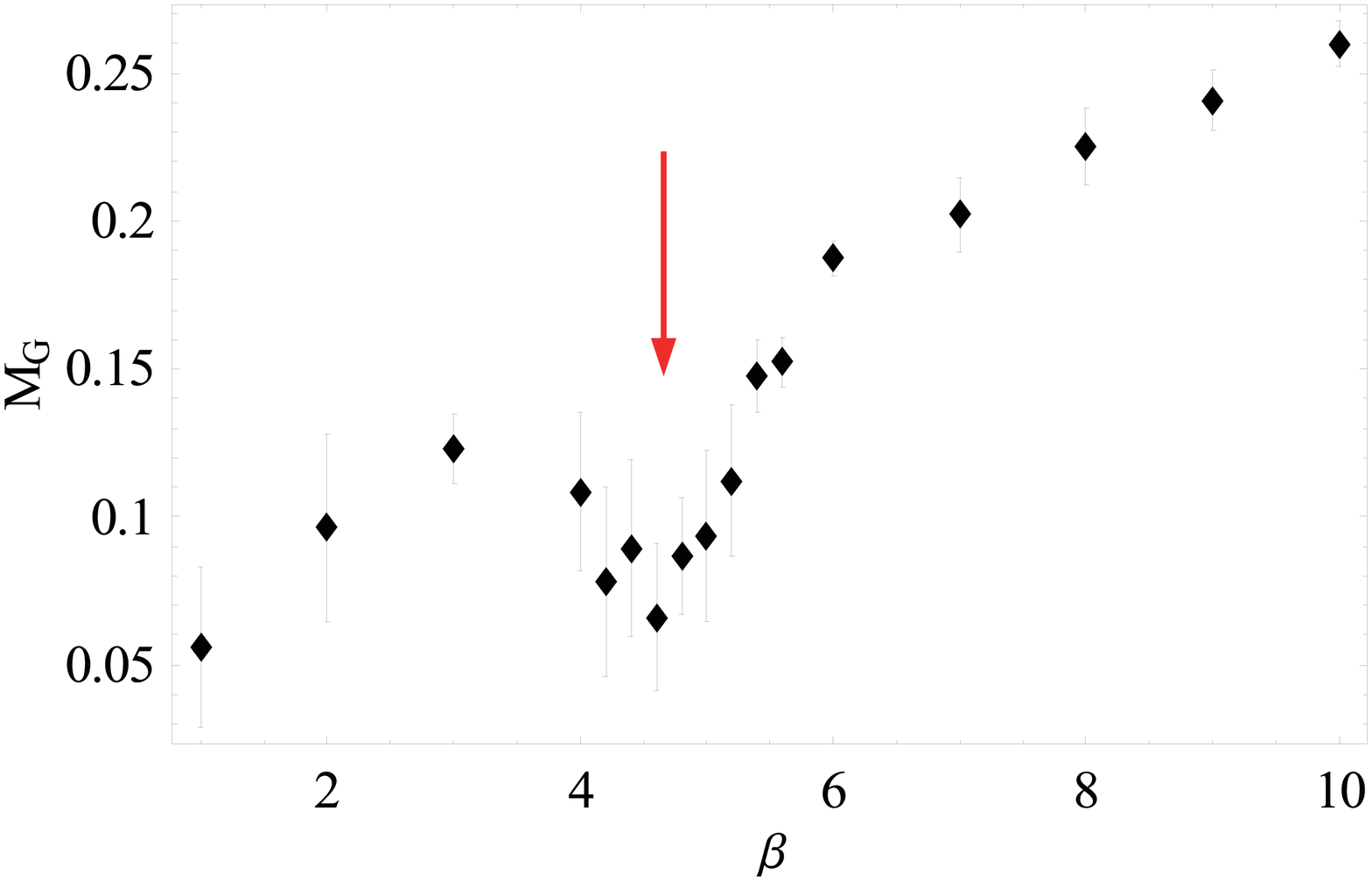}
\end{center}
\vspace{-0.7cm}
\caption{
Gauge-boson mass $M_{\rm G}$ of the external magnetic 
field propagating in $x$-$y$ plane vs $\beta$ for 
the same $c_i$ as in Fig.\ref{C:magsc}
and ${c_2}'=3.0$ ($L=16$).
At the SC phase transition point $\beta_{\rm SC}\simeq 4.5$
(indicated by an arrow) determined by the peak of $C$,
$M_{\rm G}$ starts to increase from small values. 
}
\label{MG}

\end{figure}

One of the most important phenomena of the SC is the 
appearance of a finite mass of the electromagnetic field, i.e., the Meissner effect.
In theoretical study of the SC that includes the effect of
fluctuations of the Cooper-pair wave function as in the present one, a finite mass
of the photon is the genuine order parameter of the SC.
To study it, we follow the following steps\cite{takashima};
(i) introduce a vector potential $\theta^{\rm ex}_{ri}$ for 
an external magnetic field,
(ii) couple it to Cooper pairs by replacing
$U_{ri} \to U_{ri}\exp(i\theta^{\rm ex}_{ri})$ in 
the $c_1$ term of $f_x$ 
and  add its magnetic term $f^{\rm ex}_r
=+{c_2}'(\vec{m}_r^{\rm ex})^2\ (c_2' > 0)$ to $f_r$ 
with $\vec{m}^{\rm ex}_r$  defined in the same way as (\ref{rota})
by using $\theta^{\rm ex}_{ri}$, 
(iii) let $\theta^{\rm ex}_{ri}$ fluctuate together with $z_{xa}$ and
$\theta_{ri}$ and   measure an effective mass $M_{\rm G}$
of $\theta^{\rm ex}_{ri}$ via the decay of correlation functions of
$\vec{m}_r^{\rm ex}$.
The result of $\vec{m}^{\rm ex}_r$ {\em propagating in the 1-2 plane} 
is shown in Fig.\ref{MG}.
It is obvious that the mass $M_{\rm G}$ starts to develop at
the SC phase transition point, and we conclude that Meissner effect
takes place in the SC state.

\subsection{Vortices in FMSC state}

In this section, we study the FMSC state observed in the previous section
in detail.
In particular, we are interested in whether there exist vortices and their
density if any.
There are two kinds of vortices as the present SC state contains two gaps
described by $\psi^+_r\propto z^+_r$ and 
$\psi^-_r\propto z^-_r$, where $z^{\pm}_r$ are defined as
\be
z_r^{\pm}&\equiv& {1 \over \sqrt{\mathstrut{2}}} (z_{r1}\pm iz_{r2}) 
\equiv \sqrt{\rho_x^{\pm}}
\exp(i\gamma_r^{\pm}).
\label{rhogamma}
\ee
Corresponding to the above Cooper pair fields, 
one may define the  following two kinds of 
gauge-invariant vortex
densities $V_r^+$ and $V_r^-$ in the 1-2 plane;
\be
 V_r^{\pm} &\equiv& \frac{1}{2\pi}
[{\rm mod}(\gamma_{r+1}^{\pm}-\gamma_r^{\pm}-\theta_{r1}) 
+{\rm mod}(\gamma_{r+1+2}^{\pm}-\gamma_{r+1}^{\pm}-\theta_{r+1,2}) \nn
&&\quad-{\rm mod}(\gamma_{r+1+2}^{\pm}-\gamma_{r+2}^{\pm}-\theta_{r+2,1}) 
-{\rm mod}(\gamma_{r+2}^{\pm}-\gamma_r^{\pm}-\theta_{r2})],
\ee
where mod$(x)\equiv$ mod$(x,2\pi)$. In short, $V_r^{\pm}$ 
describes vortices of electron pairs with the amplitude
$z^{\pm}_r$.

\begin{figure}[b]
\begin{center}
  \includegraphics[width=40mm]{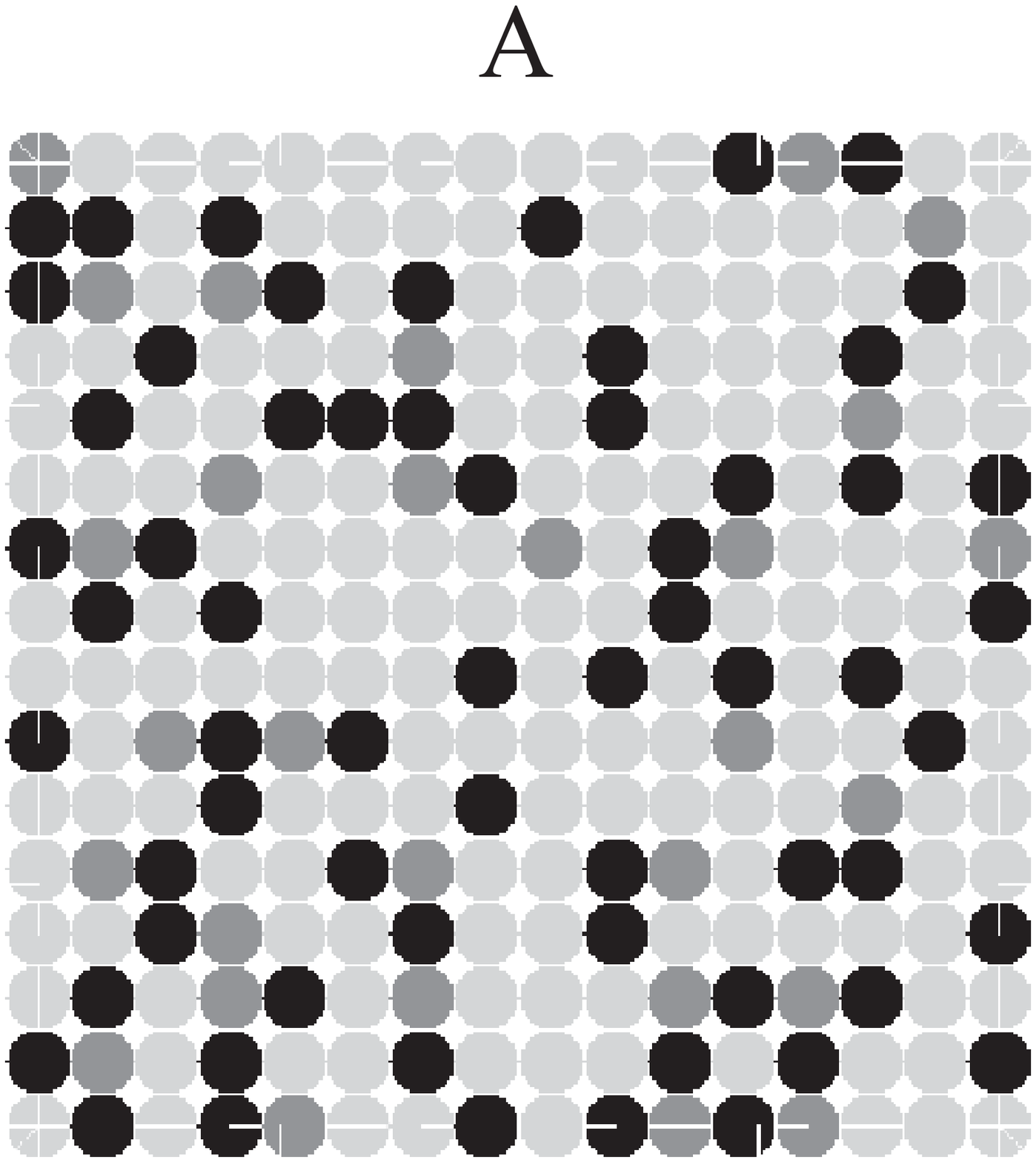}
  \includegraphics[width=40mm]{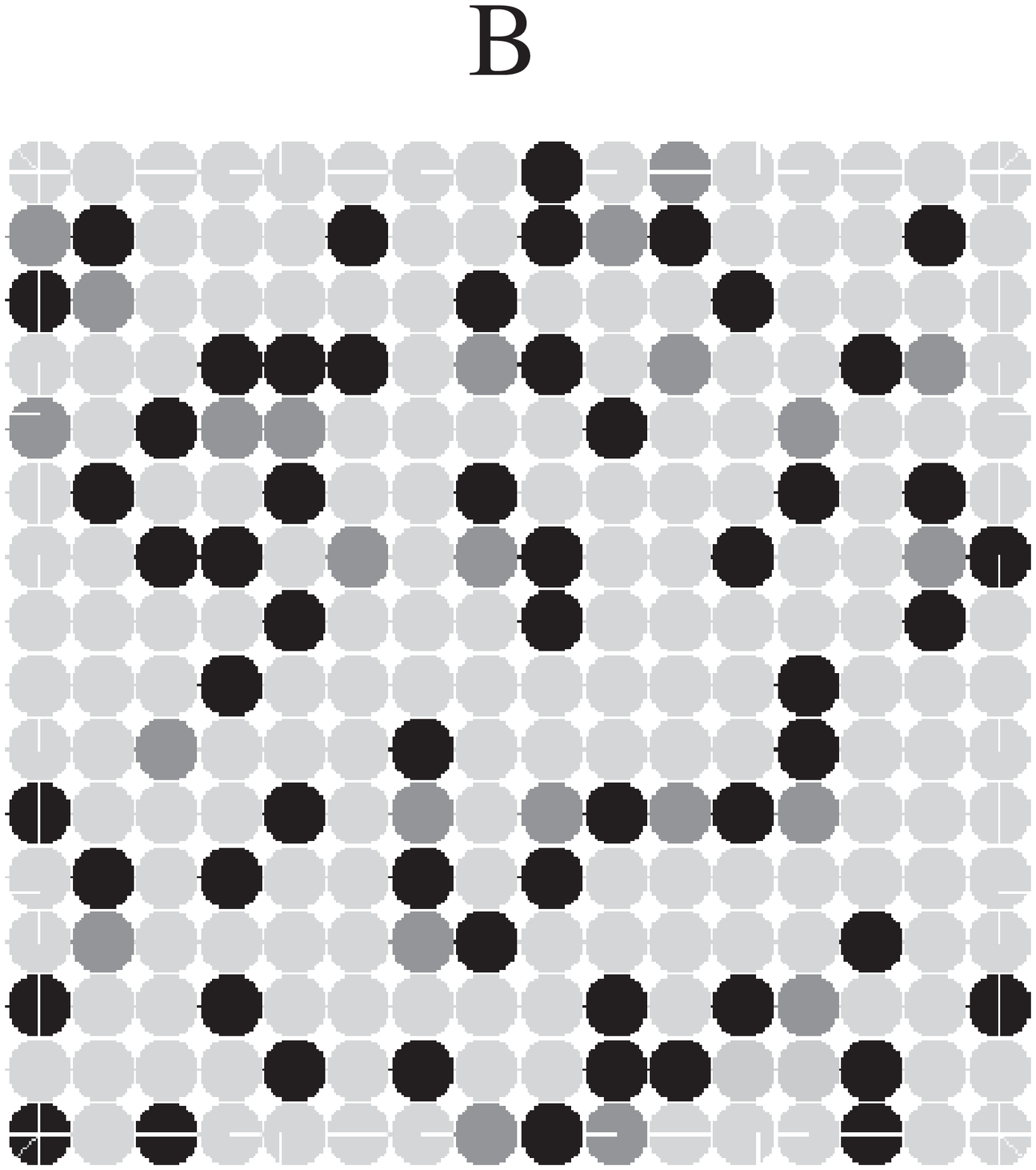}\\
\vspace{-0.7cm}
  \includegraphics[width=40mm]{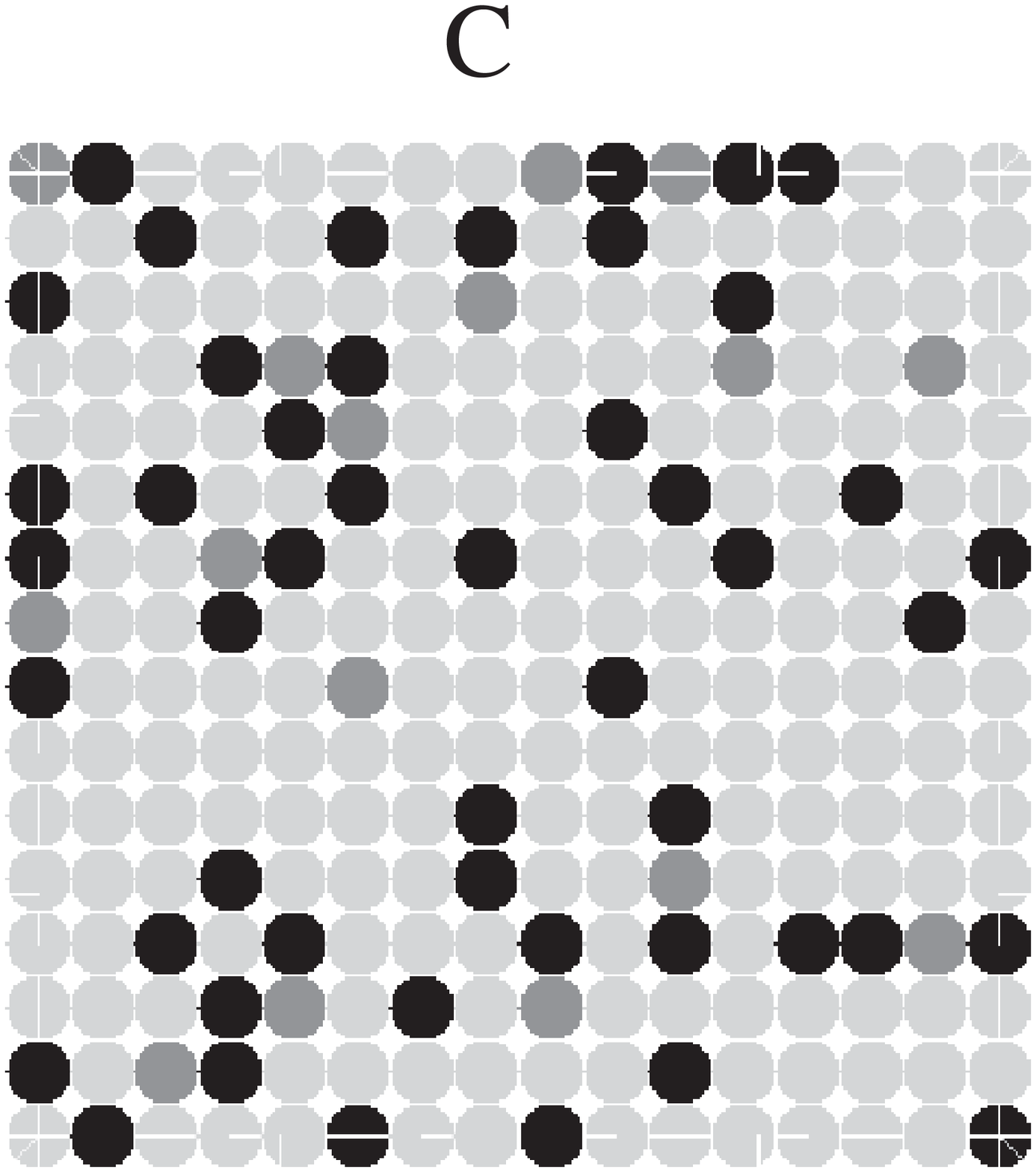}
  \includegraphics[width=40mm]{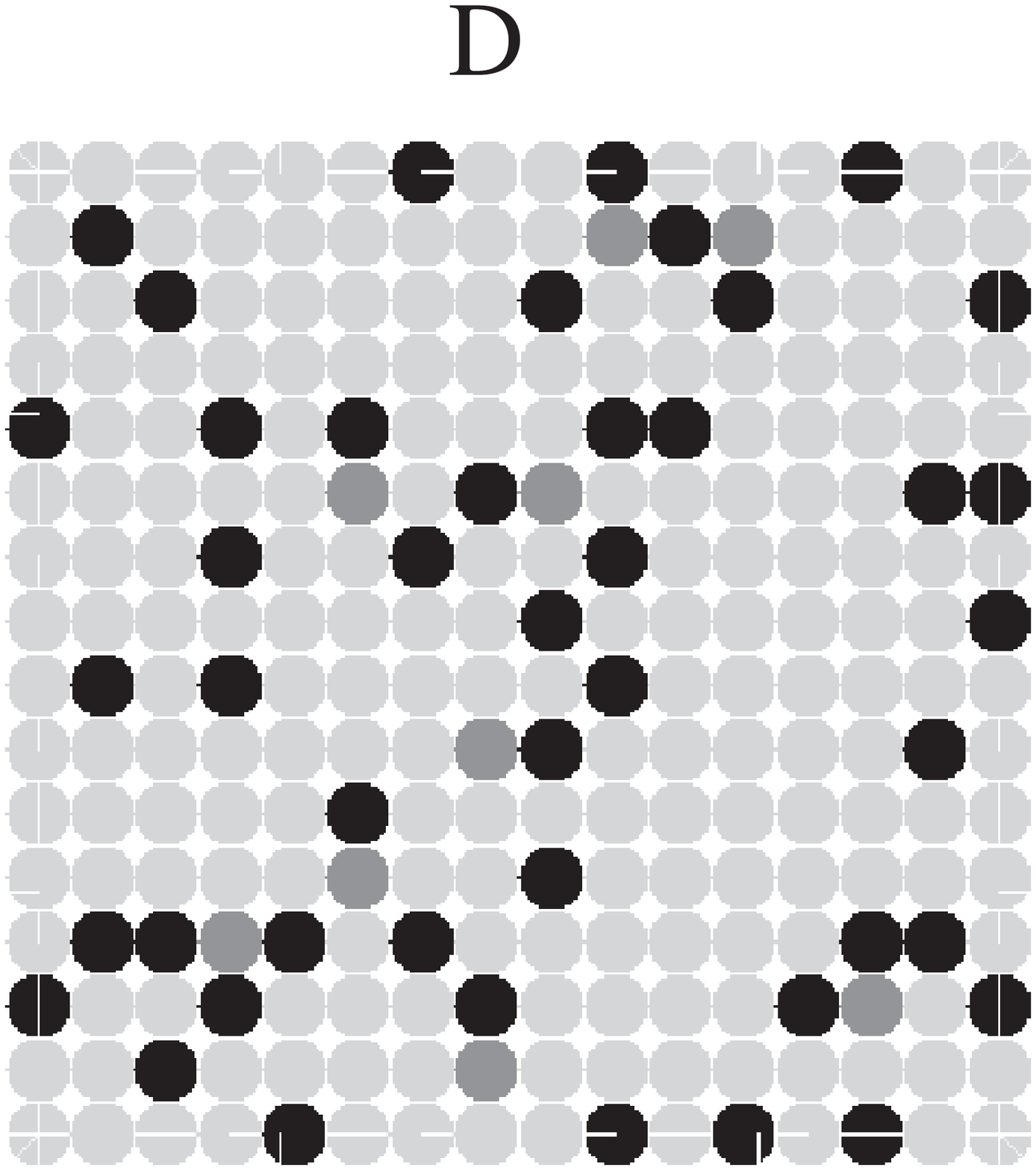}\\
\vspace{-0.3cm}
$V_r^\pm$ \\
  
\includegraphics[width=17mm]{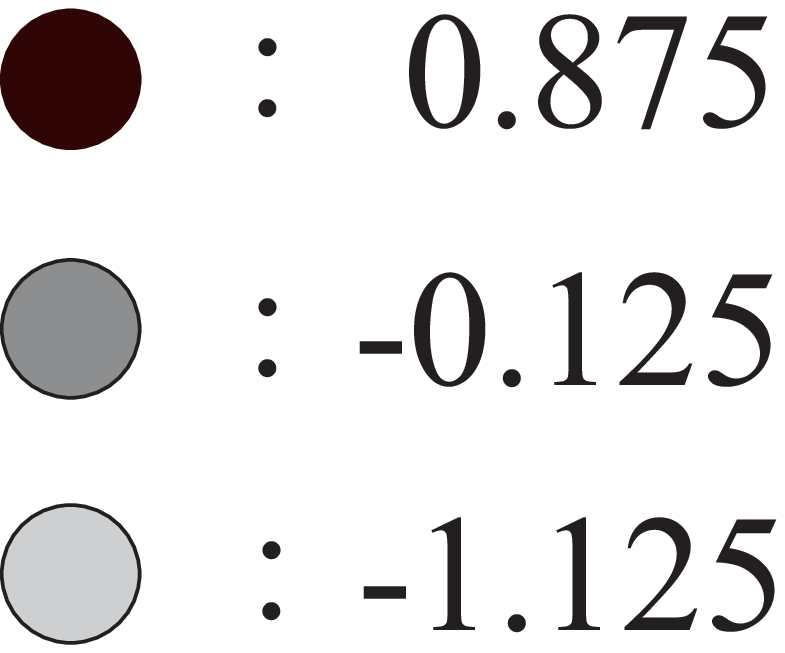}
\end{center}
\vspace{-0.3cm}
\caption{
Snapshots of vortex densities $V_r^{\pm}$ at 
$(c_1, c_2, c_3, c_4, c_5)=(0.2, 0.5, 0.2, 4.0, 1.0)$ 
for a {\it fixed magnetization} $\vec{m}_r=(0,0,\pi/4)\ (L=16)$ 
from Ref.\protect\cite{shimizu}.
Black dots; $V_r^{\pm}$= 0.875, Dark gray dots; -1.125, Light gray
dots; -0.125. 
(A)$V_r^+$ at $\beta=3.0$, (B)$V_r^-$ at $\beta=3.0$,
(C)$V_r^+$ at $\beta=7.0$, (D)$V_r^-$ at $\beta=7.0$.
The average magnitude $\la |V_{r\pm} | \ra$ is 
(A) 0.387, (B) 0.380, (C) 0.331 and (D) 0.335.
The points  $V_r^{\pm}=-0.125=-m_3/(2\pi)$ reflect $\vec{m}$ itself,
and corresponds to the state without genuine vortices.}
\label{snapvor1}  
\end{figure}

To study the behaviors of vortices step by step, 
we first consider the case of constant magnetization (magnetic field),
i.e., we set $\theta_{ri}$ by hand so that
\be
m_{r1}=m_{r2}=0,\ m_{r3}=f={\rm constant}, 
\label{constM}
\ee
freezing their fluctuations.
As the above numerical studies show that a typical magnitude of the
magnetization $\langle m_{r3}\rangle=0.8\cdots$, we put $f={\pi \over 4}$.
In this case, the SC phase transition takes place at $\beta=4.8$.
Snapshots of vortices are shown in Fig.\ref{snapvor1}.
It is obvious that at lower-$T$, i.e., $\beta=7.0$, densities of vortices 
are low compared to the higher-$T$ ($\beta=3.0$) case
but they are still nonvanishing.
This result obviously comes from the existence of the finite magnetization, 
i.e., the internal magnetic field.

It is interesting to study behavior of vortices and the magnetization 
in the FMSC state rather in detail. We now allow for fluctuations
of $\theta_{ri}$.
In this state, the magnetization $m_{r3}$ fluctuates around its mean value.
As we showed in the previous section, the FM and SC phase transitions
take place at $\beta_{\rm FM}\simeq 2.1$ and $\beta_{\rm SC}\simeq 4.5$,
respectively.
We are interested in the $\beta$-dependence of various quantities
such as the magnetic field $\la m_{r3} \ra$, the vortex densities 
$V^{\pm}_r$,
and the Cooper-pair densities $\la \rho^{\pm}_r \ra$.

\begin{figure}[t]
\begin{center}
\includegraphics[width=8cm]{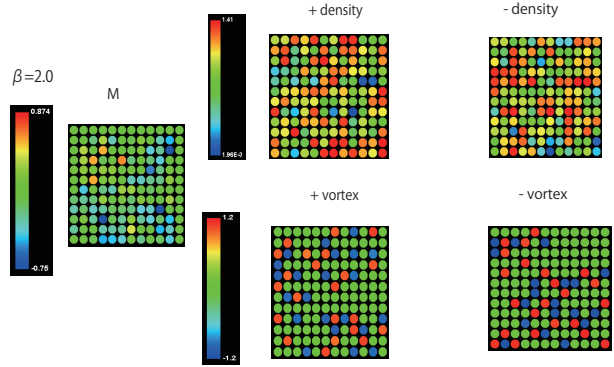}
\end{center}
\caption{
Snapshot at $\beta=2.0$ with {\it dynamically fluctuating} $\vec{m}_r$. 
The values of $c_i$ are $(c_1, c_2, c_3, c_4, c_5)=(0.2, 0.5, 0.2, 4.0, 1.0)$.
}
\label{fig6}
\label{figb20}
\end{figure}

\begin{figure}[b]
\begin{center}
\includegraphics[width=8cm]{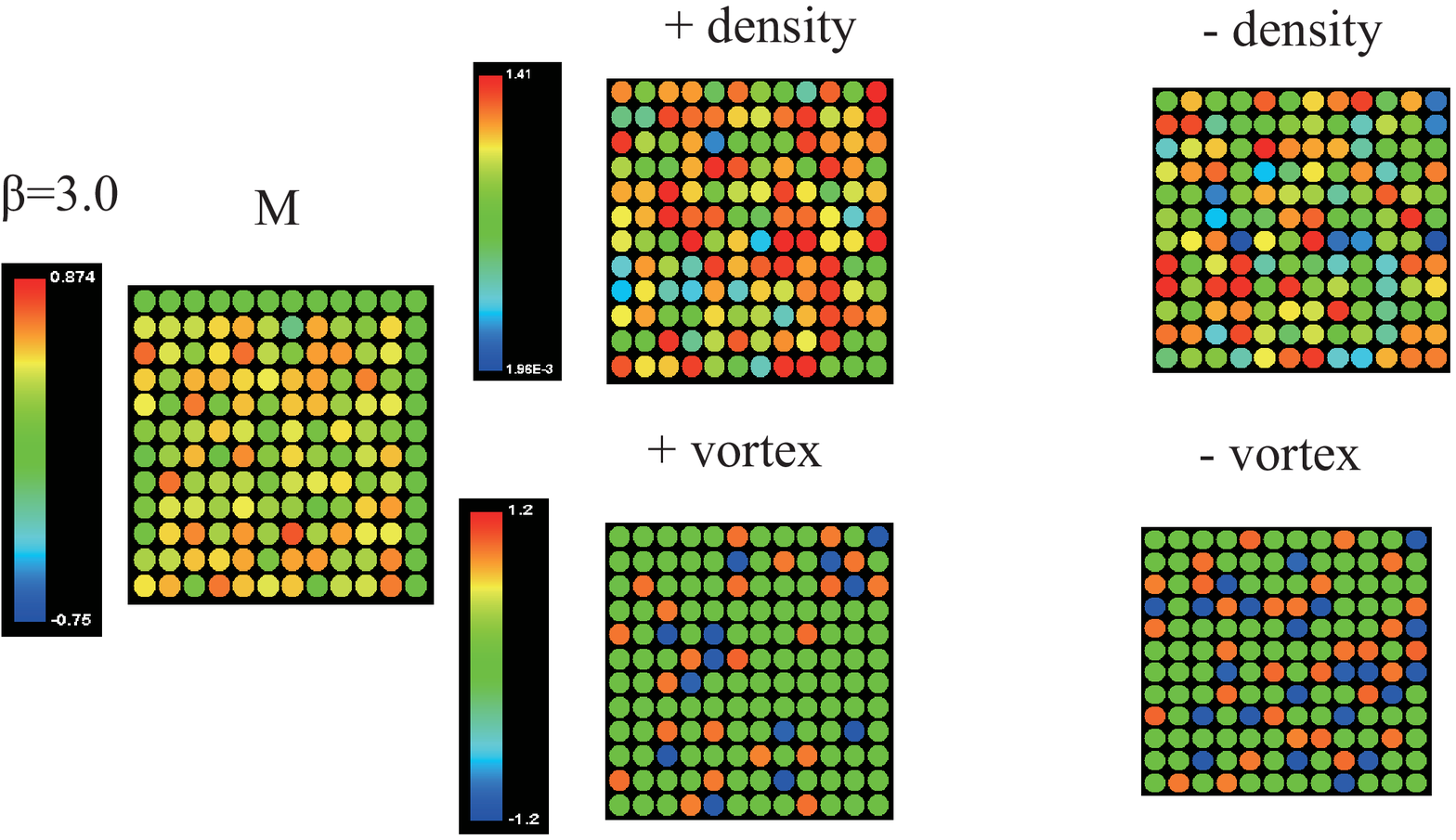}

\end{center}
\vspace{-1.7cm}
\caption{
Snapshot at $\beta=3.0$ with {\it dynamically fluctuating} $\vec{m}_r$. 
The values of $c_i$ are the same as in Fig.\ref{figb20}.
}
\label{figb30}
\vspace{-0.5cm}
\end{figure}
\begin{figure}[t]
\begin{center}
\includegraphics[width=8cm]{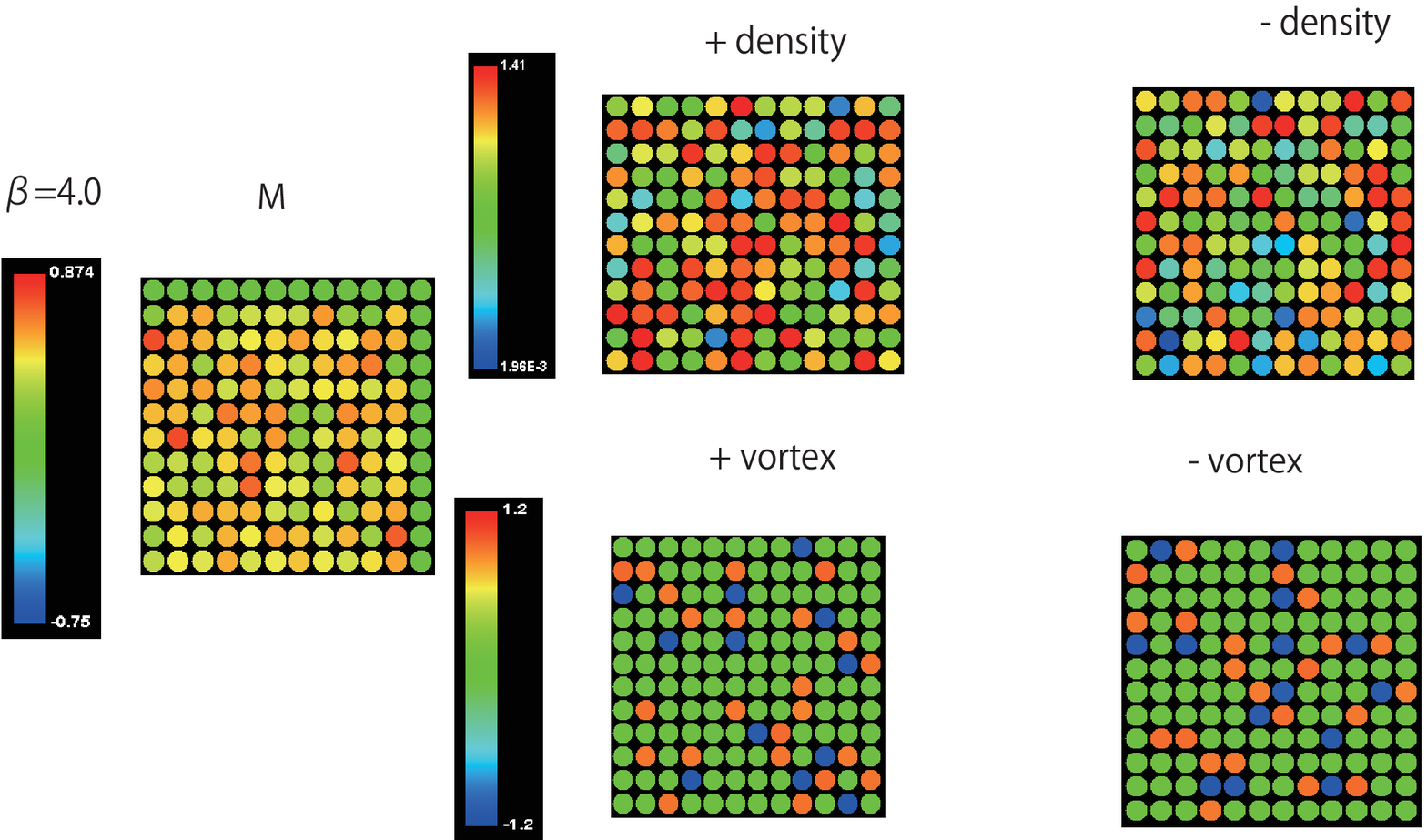}
\end{center}
\caption{
Snapshot at $\beta=4.0$ with {\it dynamically fluctuating} $\vec{m}_r$. 
The values of $c_i$ are the same as in Fig.\ref{figb20}.}
\label{figb40}
\end{figure}
\begin{figure}[b]
\begin{center}
\includegraphics[width=8cm]{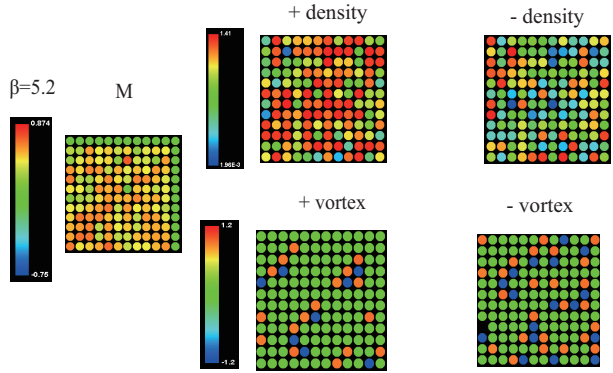}
\end{center}
\caption{
Snapshot at $\beta=5.2$ with {\it dynamically fluctuating} $\vec{m}_r$. 
The values of $c_i$ are the same as in Fig.\ref{figb20}.}
\label{figb52}
\end{figure}

In Figs. \ref{figb20} $\sim$ \ref{figb52} we show these quantities as snapshots
at $\beta= 2.0, 3.0, 4.0, 5.2$ for the set of parameters
$(c_1, c_2, c_3, c_4, c_5)=(0.2, 0.5, 0.2, 4.0, 1.0)$.
At $\beta=2.0$ (Fig.\ref{figb20}), the local magnetization $m_{r3}$ fluctuates rather 
strongly, and also the density of Cooper pairs $\rho^+_r$ and $\rho^-_r$  
are almost equal.
Densities of vortices $V^{\pm}_r$ are also fairly large.
At $\beta=3.0$(Fig.\ref{figb30}), $m_{r3}$ tends to have
a positive value.
This corresponds to the fact that the system has the FM long-range order.
At $\beta=4.0$(Fig.\ref{figb40}), $\rho^+_r$ is getting larger
compared to $\rho^-_r$.
Similarly, $V^-_r$ is sightly larger than $V^+_r$.
These results reflect the Zeeman coupling.
At $\beta=5.2$(Fig.\ref{figb52}), vortex-antivortex pairs are formed, and $V^-_r$
 is larger than $V+_r$.
As in the case of the constant magnetic field, 
the densities of vortices are nonvanishing even in the SC phase.



\section{Conclusion}
\setcounter{equation}{0}

In this paper, we present a brief review of LGT with Higgs matter field, 
and then apply it to the GL lattice model of FMSC.
The detailed MC simulations of the model reveal the global structure
such as the phase diagram 
and the physical characteristics  of each phase in an explicit manner. 
The concepts and knowledges of LGT considerably help us to build an appropriate
lattice model as well as  to interpret 
the results of MC simulations. 
We want to add this example to the  rather long list of studies 
that have utilized the following common three-step method:
(i) Establishing a lattice gauge model of the phenomenon of one's interest,
(ii) Performing MC simulations of the model,
(iii) Intepreting the MC results according to LGT.
Here we list up a couple of tips to remove obstacles in carrying out this program.
\begin{itemize}
\item
MC simulations become more effective with less variables.
To avoid unnecessary inflation of degrees of freedom,
one should identify what are  the relevant variables.
Neglecting the radial degrees of freedom of the Higgs field 
$\phi_r$ is an example\cite{radialcomp}.
\item
MC simulations require positive measure to make use of a probabilistic
(Markoff) process. Fermionic systems in their original forms 
suffer from the well known negative-sign problem,
which generates negative probabilities. By restricting
the region of parameters of the model, it could be possible to
avoid this problem. For example, the critical region around a
continuous phase transition, the order-parameters are small
and the GL expansion may assure us a positive provability\cite{lgt1}. 
\item
Another example to avoid nonpositive provability is the
case of bosonic system at finite temperatures. The path integral of
bosonic variables usually contain the imaginary kinetic term
like $i\int d\tau \bar{\phi}(d\phi/d\tau)$ in the action, 
where $\tau \in (0,\beta=1/(k_BT))$
is the imaginary time. By focusing in the high-temperature region,
the $\tau$-dependent modes of $\phi(\tau)$ may be ignored and  
so is this kinetic term. This is the case of FMSC studied in Sect. 3 and 4,
and some of the cases studied in Ref.\cite{lgt1,lgt2}.
This procedure has an extra merit that the dimension of the
effective lattice system is reduce to $d$-dimensions instead of 
$D=d+1$ dimensions due to the absence of the $x_0$-direction.
\end{itemize}

Nowadays, it is well known that a quite wide variety
of systems of cold atoms put on an optical 
lattice may serve as a quantum simulator of certain definite quantum model
with almost arbitrary values of interaction
parameters. Such a simulator is to clarify the {\it dynamical} (time-dependent)
properties of the quantum model. 
Therefore, the importance of theoretical  analysis of such models
become increased as a guide for experiments.
For example, information on the global phase structure of a quantum model
is quite useful to perform experiments in an effective manner,
although the former is confined to the static (time-independent) properties
in thermal equilibrium.
We expect that the above path  (i-iii) with the successive tips
may be useful for this purpose, in particular to simulate the model including 
compact or noncompact gauge fields. We hope as many readers as possible make use 
of this practical but powerful tool to strengthen and expand their researches.  

\vspace{1cm}
\noindent
{\bf Acknowledgments}\\

We thank Mr. T. Noguchi, our coauthor of Ref.\cite{noguchi}, for his help and 
discussions. 
We also thank Dr. K. Kasamatsu for reading the manuscript and presenting 
comments and suggestions.
This work was supported by JSPS KAKENHI Grantnumbers
23540301, 26400246, 26400412.
 


\begin{thebibliography}{0}
\bibitem{charge}
In the relativistic theory, the charge-conjugate pair of
$\phi$-particle (antiparticle) having the charge $-g$ should appear
on an equal footing.
\bibitem{cvsnc}
Sometimes the noncompact model is studied just for the purpose of
comparison with the compact model.
In general, one must ask for every system and situation under consideration 
which model (compact vs. noncompact) is more suitable.
\bibitem{uubar}
In some literatures, the term $|\phi_{x+\mu}-U_{x\mu}\phi_x|^2$
appears instead of $|\phi_{x+\mu}-\bar{U}_{x\mu}\phi_x|^2$.
They are related by $\phi_x\leftrightarrow \bar{\phi}_x$
or $U_{x\mu}\leftrightarrow \bar{U}_{x\mu}$
and equivalent each other.
\end{thebibliography}
\end{document}